%% file: main.tex
\documentclass[conference]{IEEEtran}
\IEEEoverridecommandlockouts
\usepackage{listings}
\usepackage{makeidx}
\usepackage{algorithm,algpseudocode}
\usepackage{graphicx}
\usepackage{float}
\usepackage{subcaption}
\usepackage{wrapfig}
\usepackage{amsmath}
\usepackage{rotating,todonotes, xspace}
\usepackage{multirow}
\usepackage{hyperref}
\usepackage{xspace}
\usepackage{paralist}
\usepackage{booktabs}
\newcommand{\tool}{WfCommons\xspace}
\newcommand{\rev}{\textcolor{black}}
\newcommand{\pp}[1]{\vspace{6pt}\noindent\textbf{\emph{#1 --}}\xspace}

\begin{document}

\title{\tool: A Framework for Enabling Scientific Workflow Research and Development}

\author{
  \IEEEauthorblockN{
    Tain\~a Coleman\IEEEauthorrefmark{1}\IEEEauthorrefmark{3},
    Henri Casanova\IEEEauthorrefmark{2},
    Lo\"ic Pottier\IEEEauthorrefmark{1}\IEEEauthorrefmark{3},
    Manav Kaushik\IEEEauthorrefmark{3},
    Ewa Deelman\IEEEauthorrefmark{1}\IEEEauthorrefmark{3},
    Rafael Ferreira da Silva\IEEEauthorrefmark{1}\IEEEauthorrefmark{3}
  }
  \IEEEauthorblockA{
    \IEEEauthorrefmark{1}Information Sciences Institute, University of Southern California, Marina Del Rey, CA, USA \\
    \IEEEauthorrefmark{2}Information and Computer Sciences, University of Hawaii, Honolulu, HI, USA \\
    \IEEEauthorrefmark{3}University of Southern California, Department of Computer Science, Los Angeles, CA, USA\\
    \{tcoleman,lpottier,deelman,rafsilva\}@isi.edu, henric@hawaii.edu, manavkau@usc.edu\\
  }
}

%%%%%%%%%%%%%%%%%%%%%%%%%%%%%%%%%%%%%%%%%%%%%%%%%%%%%%%%%%%%%%%%

\maketitle
\thispagestyle{empty}
\pagestyle{empty}

\begin{abstract}
Scientific workflows are a cornerstone of modern scientific computing. They
are used to describe complex computational applications that require
efficient and robust management of large volumes of data, which are
typically stored/processed on heterogeneous, distributed resources. The
workflow research and development community has employed a number of
methods for the quantitative evaluation of existing and novel workflow
algorithms and systems. In particular, a common approach is to simulate
workflow executions. In previous works, we have presented a collection of
tools that have been adopted by the community for conducting workflow
research.  Despite their popularity, \rev{they suffer from} several
shortcomings that prevent easy adoption, maintenance, and consistency with
the evolving structures and computational requirements of production
workflows. In this work, we present \emph{\tool},
a framework that provides a collection of
tools for analyzing workflow executions, \rev{for producing
generators of synthetic workflows}, and for simulating workflow
executions.  We demonstrate the realism of the generated synthetic
workflows by comparing their simulated executions to
real workflow executions.  We also contrast these results with results
obtained when using the previously available collection of tools. We find
that \rev{the workflow generators that are automatically
constructed by our framework} not only generate representative
\rev{same-scale} workflows (i.e., with structures and task
characteristics distributions that resemble those observed in
real-world workflows), \rev{but also do so at scales larger than
that of available real-world workflows}. \rev{Finally, we conduct
a case study to demonstrate the usefulness of our framework for estimating
the energy consumption of large-scale workflow executions.}
\end{abstract}


%%%%%%%%%%%%%%%%%%%%%%%%%%%%%%%%%%%%%%%%%%%%%%%%%%%%%%%%%%%%%%%%

% sections
\input{sec-introduction}
\input{sec-related-work}
\input{sec-wfcommons}
\input{sec-experiments}
\input{sec-energy}
\input{sec-conclusion}

%% acknowledgments
\section*{Acknowledgments}
This work is funded by NSF contracts \#2016610, and \#2016619;
and partly funded by DOE contract number \#DE-SC0012636, and NSF
contracts \#1923539, \#1923621, and \#1664162.
We also thank the NSF Chameleon Cloud for providing time grants
to access their resources.

% bibliography
\bibliographystyle{IEEEtran}
\bibliography{references}

% \balance

\end{document}

%% file: sec-introduction.tex
\section{Introduction}
\label{sec:introduction}

Scientific workflows are relied upon by thousands of
researchers~\cite{deelman2019evolution} for managing data analyses,
simulations, and other computations in almost every scientific
domain~\cite{liew2016scientific}. Scientific workflows have underpinned
some of the most significant discoveries of the last
decade~\cite{deelman-fgcs-2015, klimentov2015next}. These discoveries are
in part a result of decades of Workflow Management System (WMS) research,
development, and community engagement to support the
sciences~\cite{osti_1422587, ferreiradasilva2021wcs}.
As workflows continue to be adopted by
scientific projects and user communities, they are becoming more complex
and require more sophisticated workflow management capabilities. Workflows
are being designed that can analyze terabyte-scale datasets, be composed of
millions of individual tasks that execute for milliseconds up to several
hours, process data streams, and process static data in object stores.
Catering to these workflow features and demands requires WMS research
and development at several levels, from algorithms and systems all the way
to user interfaces.

A traditional approach for testing, evaluating, and evolving WMSs is to use
full-fledged software stacks to execute applications on distributed
platforms and testbeds. Although seemingly natural, this approach has
severe shortcomings including lack of reproducible results, limited
platform configurations, and time and operational costs. An alternative
that reduces these shortcomings is to use simulation, i.e., implement and
use a software artifact that models the functional and performance
behaviors of software and hardware stacks of interest. Thus, the scientific
workflow community has leveraged simulation for the development and
evaluation of, for example, novel algorithms for scheduling, resource
provisioning, and energy-efficiency, workflow data footprint constraints,
exploration of data placement strategies, among others~\cite{canon2020,
han2019, coleman2021iccs}.

Studying the execution of workflows in simulation requires sets of workflow
application instances to be used as benchmarks. This is so that
quantitative results are obtained for a range of representative workflows.
In~\cite{ferreiradasilva-escience-2014}, we have described a collection of
tools and data that together have enabled research and development of the
Pegasus WMS~\cite{deelman-fgcs-2015}, and have also \rev{been used
extensively by the workflow community}\footnote{As seen by the high number of
citations on Google Scholar.}. Despite the popularity of this pioneer
effort, it lacks (\emph{i})~a common format for representing a workflow's
execution in a way that is \rev{agnostic to WMS that was used to execute it};
(\emph{ii})~\rev{methods for deriving workflow structure and workflow task
performance characteristics based on workflow execution}; and
(\emph{iii})~techniques for \rev{automating the process of producing
generators of realistic synthetic workflows for any given workflow application}.

In this paper, we present \emph{\textbf{\tool}}~\cite{wfcommons}, an open
source framework that provides a collection of methods and
techniques, implemented as part of usable tools, \rev{that addresses the
shortcomings of our previous set of tools.} \tool uses a
\rev{system-agnostic} JSON format for representing workflow instances
\rev{based on execution logs}.  \tool also provides an open-source Python
package to analyze workflow instances and \rev{produce generators of
realistic synthetic workflow instances}, generated in that same format.
Workflow simulators that support this format can then take real-world and
synthetic workflow instances as input for driving the simulation.
Figure~\ref{fig:concept} shows an overview of the \tool conceptual
architecture. Information  in workflow execution logs is extracted as
\textbf{workflow instances} represented using the common JSON format.
Workflow ``recipes" are obtained from the analysis of \rev{sets of workflow
instances for a particular application}.  More precisely, a recipe embodies
results from statistical analysis and distribution fitting performed for
each workflow task type so as to characterize task runtime and input/output
data sizes.  \rev{The recipes also incorporates information regarding the
graph structure of the workflows (tasks dependencies and frequency of
occurrences), which are automatically derived from the analysis of the
workflow instances.} Each recipe is then used for automatically producing a
\textbf{workflow generator}, which in turn produces synthetic workflow
instances \rev{that are representative of the application domain}.
Finally, these instances can be used by a \textbf{workflow simulator} for
conducting experimental workflow research and development.  Specifically,
this work makes the following contributions~\footnote{A short version
of this work was previously published in~\cite{da2020workflowhub}.}:

\begin{figure}[!t]
  \centering
  \includegraphics[width=\linewidth]{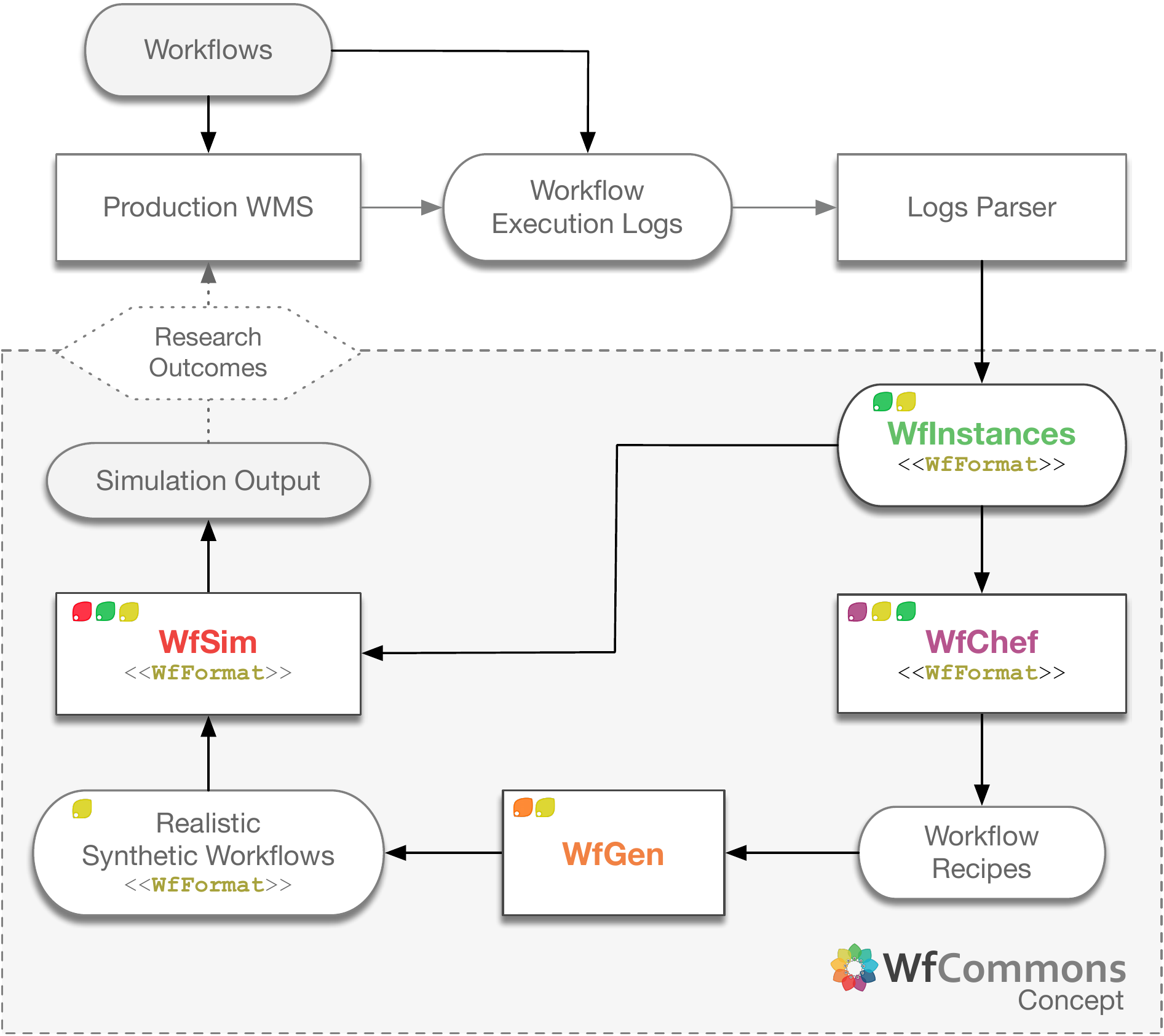}
  \caption{\rev{The \tool conceptual architecture.}}
  \label{fig:concept}
\end{figure}

\begin{compactenum}
  \item A collection of workflow execution instances acquired from actual
        executions of state-of-the-art compute- and data-intensive workflows
        in a cloud environment;
  \item A common format for representing both collected instances and
        generated synthetic instances;
  \item An open source Python package~\cite{wfcommons-python} that provides
        methods for analyzing instances, deriving recipes, and generating
        representative synthetic instances;
  \item A collection of open-source WMS simulators
        and simulation frameworks that support our common format;
  \item \rev{An evaluation of the accuracy of \tool' generated synthetic workflows
        and a comparison to our previous
        sets of tools~\cite{ferreiradasilva-escience-2014, da2020workflowhub};}
  \item \rev{A case study in which we demonstrate the usefulness of \tool in the
         context of energy-efficient workflow executions.}
\end{compactenum}

This paper is organized as follows. Section~\ref{sec:related-work}
discusses related work. The \tool project and the associated concepts and
tools are explained in Section~\ref{sec:wfcommons}.
Section~\ref{sec:experiments} provides an experimental evaluation of the
accuracy of \tool' generated synthetic workflows.
Section~\ref{sec:energy} presents our case-study on energy-efficient
workflow executions.  Finally, Section~\ref{sec:conclusion} concludes with
a summary of results and perspectives on future work.

%% file: sec-related-work.tex
\section{Related Work}
\label{sec:related-work}

Workload archives are widely used for distributed computing research, to
validate assumptions, to derive workload models, and to perform experiments
by replaying workload executions, either in simulation or on real-world platforms.
Available repositories, such as the Parallel Workloads
Archive~\cite{feitelson2014experience}, the Grid Workloads
Archive~\cite{iosup2008grid}, and the Failure Trace
Archive~\cite{kondo2010failure}, contain data acquired at the
infrastructure level at compute sites, or by monitoring or obtaining logs
from deployed compute services. The workloads in these archives do include
data generated by workflow executions. However, the information captured is
about individual job executions and about resource utilization. As
a result, there is at best little information on the task dependency
structure of workflows.

In the context of scientific workflows,
% the myExperiment
% project~\cite{goble2010myexperiment} enables users to share workflows and
% their semantics, however it does not collect metrics from
% workflow executions.
the Common Workflow Language
(CWL)~\cite{amstutz2016common} is an open standard for describing workflows
in a way that makes them portable and scalable across a variety
of software and hardware environments. Our proposed common format
(described below) is conceptually similar to the CWL standard, though our format
captures performance metrics data (e.g., volumes of I/O reads and
writes, runtime, power consumption, etc.) and compute resource characteristics,
which are key for generating realistic workflow instances. The recently
established Workflow Trace Archive~\cite{versluis2020workflow} is an
open-access archive that provides a collection of execution instances from
diverse computing infrastructures and tools for parsing, validating, and
analyzing instances. To date, the archive has collected instances from 11
existing online repositories (including 10 instances obtained
from a preliminary version of WfCommons) and uses an object-oriented
representation (based on the Parquet columnar storage format used in Hadoop) for
documenting instances. Our format instead uses JSON, which is agnostic to
the programming language used for processing instances.
Also, the format used in~\cite{versluis2020workflow} captures workflow
executions information in terms of resource usage on the specific hardware
platform used to execute the workflow. As a result, it is difficult to use
this information to reconstruct a platform-independent, abstract workflow
structure.  By contrast, while WfCommons also records platform-specific
behaviors in its instances, in addition it ensures that the abstract workflow
structure is directly available from these instances.  This is crucial for
research purposes, as abstract workflow structures are needed for,
for instance, simulating workflow executions  on platform configurations that
differ from that used to collect the workflow execution instance.

Several studies have used synthetic workflows to explore how different
workflow features impact execution and interplay with each other (e.g.,
number of tasks, task dependency structure, task execution times).  Tools
such as SDAG~\cite{amer2012evaluating} and \rev{DAGGEN~\cite{daggen} generate
random workflow instances based on the number of tasks,
the maximum number of levels that can be spanned by an edge, the edge density,
the data-to-computation ratio, the width, etc. DAGEN~\cite{amalarethinam2011dagen}
generates random DAGs for parallel programs modeled according real instances
of parallel programs with respect to task computation and communication
payloads. DAGITIZER~\cite{amalarethinam2012dagitizer}, an extension of
DAGEN-A, is applicable to grid workflows in which all parameters are
generated randomly. Although these generators can produce a very diverse
set of DAGs, they may not necessarily be representative of real-world
scientific workflows.}

\rev{Using the structure of real-world workflow instances to generate DAGs
for specific applications is an alternative to the random generation approach.
The work in~\cite{garijo2014common} identifies workflow ``motifs", or
sub-structures, that are used to reverse engineer workflow structures based
on the data created and used by the tasks. Although these motifs allow for
automated workflow generation, identifying them is an arduous manual
process. The work in~\cite{van2003workflow} targets business process
requirements such as parallelism, choice, synchronization, etc., and
identifies over forty workflow patterns. These patterns can be mapped to
structures in real scientific workflows~\cite{yildiz2009towards}, but they
do not necessarily respect the ratios of the different types of tasks. The
problem lies in the fact that a workflow structure is not only defined by a
set of vertices and edges, but also by the task type -- i.e. executable
name, of each vertex.}

In~\cite{katz2016application}, application skeletons are used to build
synthetic workflows that represent real applications for benchmarking.
In our previous work~\cite{ferreiradasilva-escience-2014}, we developed a
tool for generating synthetic workflow configurations based on real-world
workflow  instances. As a result, the overall structure of generated
workflows was reasonably representative of real-world workflows. But that
tool uses only two types of statistical distributions  (uniform and
normal), and as a result workflow performance behavior may not be
representative (see results in Section~\ref{sec:experiments}).

%% file: sec-wfcommons.tex
\section{The \tool Framework}
\label{sec:wfcommons}

The \rev{\tool project (\url{https://wfcommons.org}) is an open source}
framework for enabling scientific workflow research and development. It
provides foundational tools for analyzing workflow execution instances, and
generating synthetic, yet realistic, workflow instances. These instances can then
be used for experimental evaluation and development of novel algorithms
and systems for overcoming the challenge of achieving efficient and robust
execution of ever-demanding workflows on increasingly complex distributed
infrastructures.

Fig.~\ref{fig:concept} shows an overview of the research \rev{and development} life cycle
that integrates the \rev{four major components \tool:
(i)~workflow execution instances (WfInstances), (ii)~workflow recipes (WfChef),
(iii)~workflow generator (WfGen), and (iv)~workflow simulator (WfSim).}

\rev{\subsection{WfInstances}}
\label{sec:instances}

\begin{table*}[!t]
  \centering
  \scriptsize
  \caption{Collection of workflow execution instances hosted in WfInstances. All instances were obtained using the Pegasus \rev{and Makeflow WMSs} running on the Chameleon cloud platform.}
  \label{tab:instances}
  \begin{tabular}{llllrrp{7.6cm}}
    \toprule
    Application & WMS & Science Domain & Category & \# instances & \# Tasks & Runtime and Input/Output Data Sizes Distributions \\
    \midrule
    1000Genome  & Pegasus & Bioinformatics  & Data-intensive & 22 &  8,844
                & alpha, chi2, fisk, levy, skewnorm, trapz \\
    \rev{BLAST} & \rev{Makeflow} & \rev{Bioinformatics} & \rev{Compute-intensive} & \rev{15} & \rev{2,245} & \rev{arcsine, argus, trapz} \\
    \rev{BWA}   & \rev{Makeflow} & \rev{Bioinformatics} & \rev{Data-intensive} & \rev{15} & \rev{10,560} & \rev{arcsine, argus, rdist, trapz} \\
    Cycles      & Pegasus & Agroecosystem   & Compute-intensive & 24 & 30,720
                & alpha, beta, chi, chi2, cosine, fisk, levy, pareto,
                  rdist, skewnorm, triang \\
    Epigenomics & Pegasus & Bioinformatics  & Data-intensive & 26 & 15,242
                & alpha, beta, chi2, fisk, levy, trapz, wald \\
    Montage     & Pegasus & Astronomy & Compute-intensive &  \rev{17} & \rev{37,619}
                & alpha, beta, chi, chi2, cosine, fisk, levy, pareto,
                  rdist, skewnorm, wald \\
    Seismology  & Pegasus & Seismology & Data-intensive & 11 &  6,611
                & alpha, argus, fisk, levy \\
    SoyKB       & Pegasus & Bioinformatics & Data-intensive & 10 &  3,360
                & argus, dweibull, fisk, gamma, levy, rayleigh, skewnorm,
                  triang, trapz, uniform \\
    \rev{SRA Search} & \rev{Pegasus} & \rev{Bioinformatics} & \rev{Data-intensive} & \rev{25} & \rev{1,580} & \rev{arcsine, argus, beta, dgamma, fisk, norm, rdist, trapz}\\
    \midrule
    \rev{9 applications} & 2 WMSs & 4 domains & 2 categories & \rev{165} & \rev{116,781} & \rev{21} probability distributions \\
    \bottomrule
  \end{tabular}
\end{table*}

\rev{Catalogs of workflow instances are instrumental for
evaluating workflow solutions in simulation or in real
conditions. The WfInstances component} targets the collection and curation
of open-access production workflow instances from various
scientific applications, all made available using a common format.
A workflow instance is built based on logs of an actual execution
of a scientific workflow on a distributed platform (e.g., clouds, grids,
clusters) \rev{using a WMS}.  Specifically, the three main types of
information included in the instance are:

\begin{compactitem}
  \item Workflow task execution metrics (runtime, input and output data sizes,
        memory used, energy consumed, CPU utilization, compute resource that
        was used to execute the task, etc.);
  \item Workflow structure information (inter-task control and data
        dependencies); and
  \item Compute resource characteristics (CPU speed, available RAM, etc.).
\end{compactitem}

\pp{Workflow Instance Format}
The \tool project uses a common format, \rev{{\bf WfFormat}}, for
representing collected workflow instances and generated synthetic workflows
instances. Workflow simulators and simulation frameworks that support
WfFormat can then use both types of instances interchangeably. WfFormat
uses a JSON specification (available on
GitHub~\cite{wfcommons-schema}), which captures all relevant instance
information as listed above.  The GitHub repository also provides a
Python-based JSON schema validator for verifying the syntax of JSON
instance files, as well as their semantics, e.g., whether all files and
task dependencies are consistent. Users are encouraged to contribute
additional workflow instances for any scientific domain, as long as they
conform to WfFormat.  \rev{Currently, \tool provides parsers for
converting execution logs into WfFormat for two state-of-the-art WMSs:
Pegasus~\cite{deelman-fgcs-2015} and Makeflow~\cite{albrecht2012makeflow}.}

\pp{Collected Workflow Instances}
An integral objective of the \tool project is to collect and
reference open access workflow instances from production workflow systems.
Table~\ref{tab:instances} summarizes the set of workflow instances currently
hosted in WfInstances. These instances are from \rev{nine} representative science
domain applications \rev{supported by Pegasus or Makeflow},
for workflows composed of compute- and/or
data-intensive tasks. (Note that although a workflow may be categorized
overall as, for example, data-intensive, it may include
CPU-intensive tasks.)  We argue that the
\rev{165} archived workflow instances form a representative set of small- and
large-scale workflow configurations. In addition to consuming/producing large
volumes of data processed by thousands of compute tasks, the structures of
these workflows are sufficiently complex and heterogeneous to encompass
current and emerging large-scale workflow execution
patterns~\cite{ferreiradasilva-fgcs-2017}.

\subsection{\rev{WfChef}}
\label{sec:wfchef}

\rev{WfChef is the \tool component that automates the construction of
synthetic workflow generators for any given workflow application. The input
to this component is a set of real workflow instances described in the
\emph{WfFormat} (e.g., instances available in \emph{WfInstances}).
WfChef automatically analyzes the real workflow instances for
two purposes. First, it discovers workflow subgraphs that represent
fundamental task dependency patterns. Second, it derives
statistical models of the workflow tasks' performance characteristics.
WfChef then outputs a ``recipe", that is, a data structure that
encodes the discovered pattern occurrences as well as the statistical models
of workflow task characteristics.
This recipe is then used by \emph{WfGen} (see Section~\ref{sec:wfgen}) to
generate realistic synthetic workflow instances with any arbitrary number of
tasks. The way in which WfChef operates is depicted in the top part
of Figure~\ref{fig:generation}, and hereafter we provide a brief
overview of the methods used to construct
recipes.  A detailed description of the algorithms and an evaluation
of their accuracy is provided in~\cite{coleman2021wfchef}.}

\pp{\rev{Finding Pattern Occurrences}}
\rev{Given a set of real workflow instances for an application, WfChef
finds the patterns that have more than one occurrence in the same workflow
graphs and records these pattern occurrences. To identify patterns, WfChef
defines a task's \textbf{type} as the kind of computation performed by the
task, which is given by the task's name as extracted from the workflow
execution logs (e.g., the name of the executable that is invoked to perform
the computation). WfChef then recursively computes a unique ID for each
task, called the \textbf{type hash}, based on the task's type and the type
hashes of all the task's ancestors and descendants. The type hash of a task
thus encodes information about the graph's structure and the role of the
task in that structure. Finally, the type hash of a sub-graph is defined as
the set of the type hashes of the tasks in that sub-graph.  WfChef defines
a (repeating) pattern as a set of disjoint sub-graphs that occur in a
workflow instance's graph that have all the same type hash (i.e., a pattern
is an equivalence class of the set of sub-graphs, where equivalence is
defined as type hash equality). Two sub-graphs with the same type hash,
i.e., with tasks of the same types and same dependency structures, do not
necessarily have the same number of tasks but are
occurrences of the same pattern.}

\rev{To find pattern occurrences, WfChef proceeds as follows:}

\rev{
\begin{compactenum}
  \item Pick two distinct tasks, $t_1$ and $t_2$, with \emph{the same type hash};
  \item $S_1=\{t_1\}$ and $S_2=\{t_2\}$.
  \item Add to $S_1$, resp. $S_2$, all the parents and children of tasks in $S_1$, reps. $S_2$;
  \item $S_1 = S_1 - S_1 \cap S_2$; $S_2 = S_2 - S_1 \cap S_2$;
  \item If $S_1$ or $S_2$ has increased in size after steps 3 and 4, go back to step 3;
  \item Otherwise, $S_1$ and $S_2$ are identified as two occurrences of the same pattern.
\end{compactenum}
}

\noindent
\rev{Note that because $t_1$ and $t_2$ have the same type hash, it is guaranteed
that $S_1$ and $S_2$ will have the same type hash, and thus are occurrences
of the same pattern.}

\begin{figure}[!ht]
  \centering
  \includegraphics[width=0.7\linewidth]{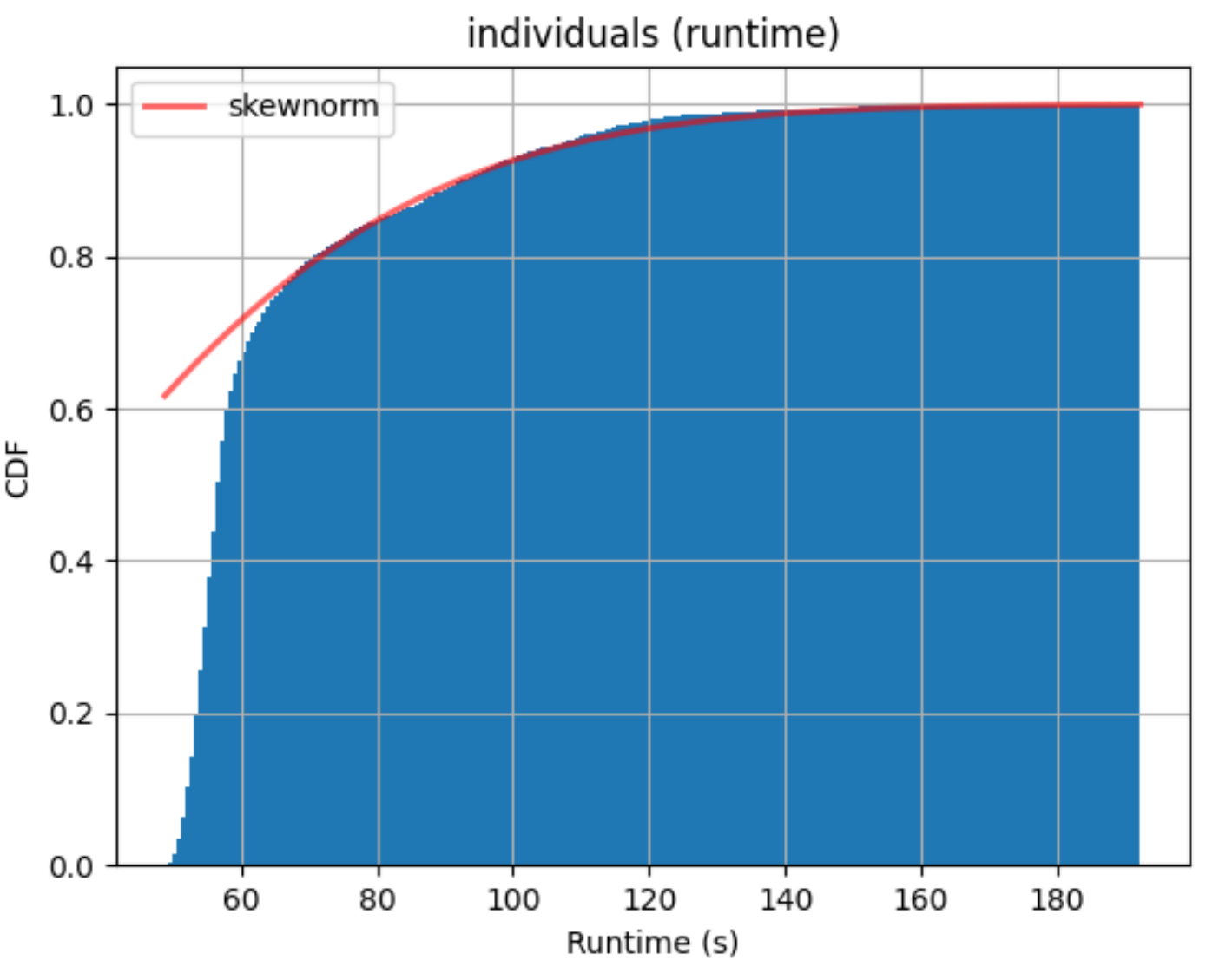} \\
  \includegraphics[width=0.7\linewidth]{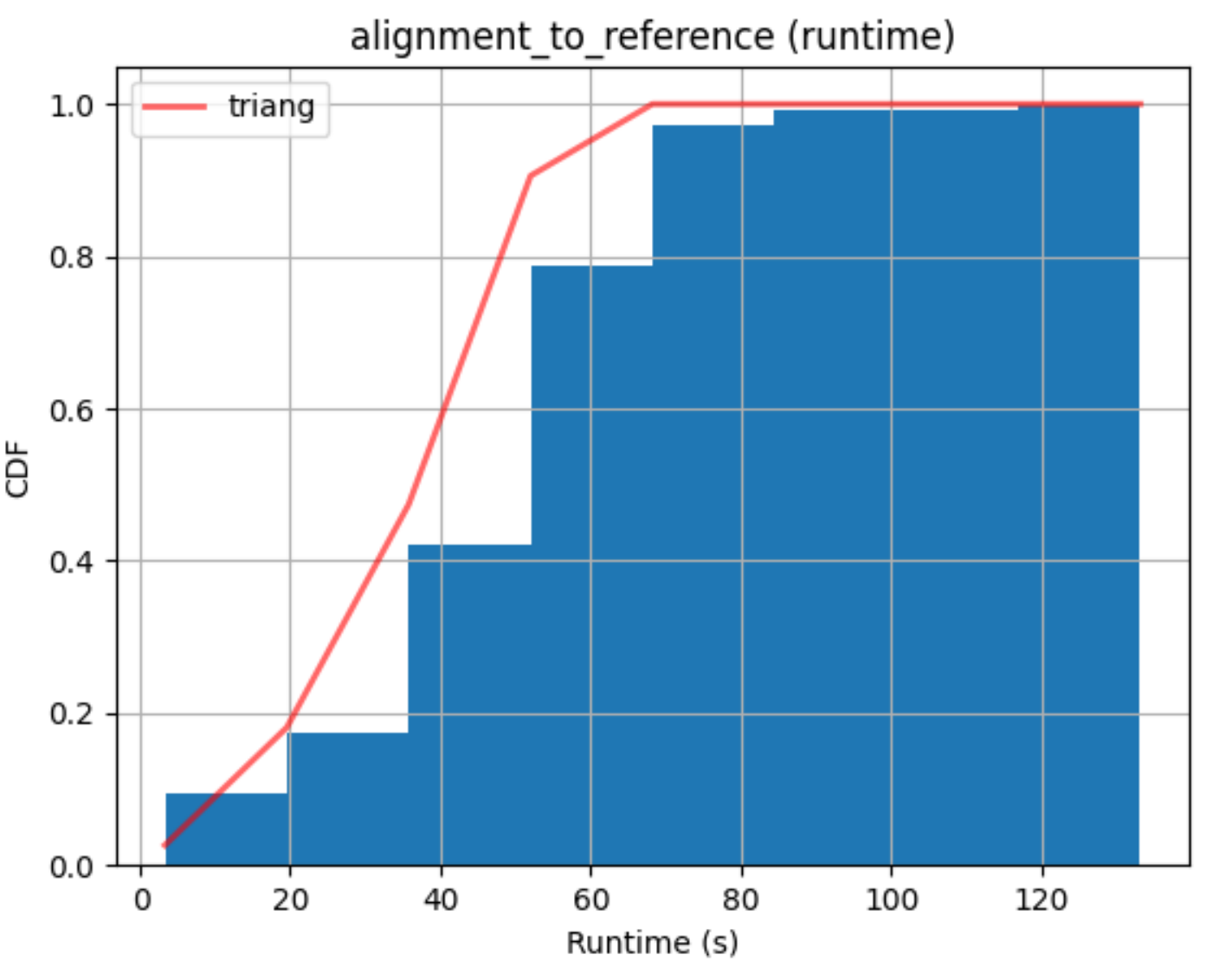}
  \caption{Example of probability distribution fitting of runtime
           (in seconds) for \texttt{individuals} tasks in 1000Genome workflows
           (\emph{top}) and \texttt{alignment\_to\_reference} tasks in
           SoyKB (\emph{bottom}) workflows.}
  \label{fig:fitting-runtime}
\end{figure}

\pp{\rev{Modeling Performance Characteristics}}
\rev{Besides pattern occurrences, a workflow recipe also includes
statistical characterizations of task performance metrics. These are
necessary for generating representative workflow task instances (by
sampling task runtime and input/output data sizes from appropriate
probability distributions).  Specifically, WfChef analyzes a set of real
workflow instances for a particular application to produce statistical
summaries of workflow performance characteristics, per task type. To this
end, WfChef performs probability distributions fitting (minimizing the mean
square error).} Figure~\ref{fig:fitting-runtime} shows an example of
probability distribution fitting of task runtime for two task types from
different workflow instances, by plotting the cumulative distribution
function (CDF) of the data and the best probability distribution found.
The outcome of this analysis applied to a set of workflow instances
\rev{for a particular application} is a summary that includes, for each
task type, the best probability distribution fits for runtime, input data
size, and output data size. For instance, Table~\ref{tab:instances} lists
(for each workflow application for which WfInstances hosts instances) the
probability distributions used for these fits. Listing~\ref{lst:recipe}
shows the statistical summary for one particular task type in the
1000Genome workflow application.

\definecolor{codegray}{rgb}{0.5,0.5,0.5}
\definecolor{codepurple}{rgb}{1,0,0}
\definecolor{backcolour}{rgb}{0.95,0.95,0.92}

\lstdefinestyle{mystyle}{
    backgroundcolor=\color{backcolour},
    numberstyle=\tiny\color{codegray},
    stringstyle=\color{codepurple},
    basicstyle=\ttfamily\scriptsize,
    breakatwhitespace=false,
    breaklines=false,
    captionpos=b,
    keepspaces=true,
    numbers=left,
    numbersep=1pt,
    showtabs=false,
    tabsize=2,
    columns=fullflexible,
    framexrightmargin=0em,
    float=tp,
    floatplacement=tbp,
}

\begin{lstlisting}[language=Python, style=mystyle, label={lst:recipe}, caption=Example of an analysis summary showing the best fit probability distribution for runtime of the \texttt{individuals} tasks in 1000Genome workflows.]
"individuals": {
    "runtime": {
        "min": 48.846,
        "max": 192.232,
        "distribution": {
            "name": "skewnorm",
            "params": [
                11115267.652937062,
                -2.9628504044929433e-05,
                56.03957070238482
            ]
        }
    },
    ...
}
\end{lstlisting}

\subsection{\rev{WfGen}}
\label{sec:wfgen}

Workflow instances are commonly used to drive experiments for
evaluating novel workflow algorithms and systems. It is crucial to run
large numbers of such experiments for many different  workflow
configurations, so as to ensure generality of obtained results.  In
addition, it is useful to conduct experiments while varying one or more
characteristics of the workflow application, so as to study how these
characteristics impact workflow execution. For instance, one may wish, for a
particular overall workflow structure, to study how the workflow execution
scales as the number of tasks increases.  And yet, current archives only
include instances for limited workflow configurations. And even as efforts are
underway, including \tool, to increase the size of these archives, it
is not realistic to expect them to include all relevant workflow
configurations for all experimental endeavors. Instead, tools must be
provided to generate representative \emph{synthetic} workflow instances. These
instances should be generated based on real workflow instances, so as to be
representative, while conforming to user-specified characteristics, so as to be useful.
\rev{The WfGen component} targets the generation of such realistic synthetic
workflow instances.

\begin{figure}[!t]
  \centering
  \includegraphics[width=0.7\linewidth]{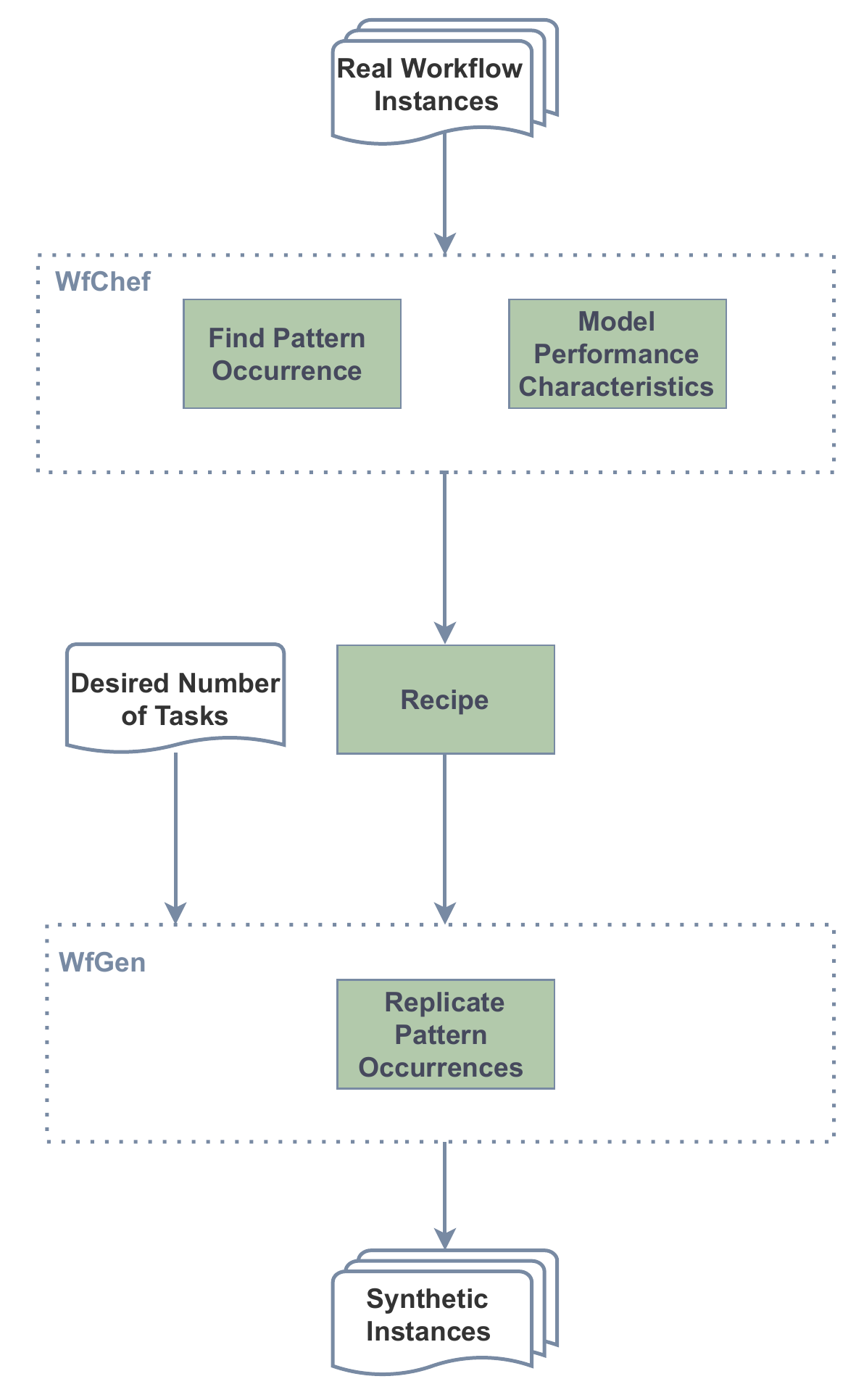}
  \caption{\rev{Overview of the synthetic workflow instance generation process.
  WfChef starts by analyzing a set of real workflow instances to discover
  pattern occurrences and compute statistical summaries of workflow task
  characteristics.  It outputs a recipe that records this information.
  WfGen takes as input a recipe and a desired number of tasks, and replicates
  pattern occurrences in real-world instances to generate a synthetic instance
  with (approximately) the desired number of tasks.}}
  \label{fig:generation}
\end{figure}

\pp{\rev{Generating Synthetic Instances}}
\rev{WfGen takes as input a \emph{workflow recipe} produced by WfChef for a
particular application and a desired number of tasks.  Note that each
workflow recipe specifies a lower bound on the number of tasks that a
generated synthetic workflow instance may contain. This is to ensure that
generated instances contain required application-specific structure.  WfGen then
automatically generates synthetic, yet realistic,  randomized workflow
instances with (approximately) the desired number of tasks. The way in
which WfGen operates is depicted in the bottom part of
Figure~\ref{fig:generation}.  To reach the designed number of tasks WfGen
iteratively \emph{replicates} pattern occurrences that are listed in the
workflow recipes, randomly picking which pattern occurrence to replicate
using a uniform probability distribution.  Replicating a pattern occurrence
simply consists in making a copy of all tasks and inter-task edges in the
pattern occurrence to generate a new sub-graph.  The entry and exit tasks
of this sub-graph are connect to the same parent and children tasks as the
entry and exit tasks of the original pattern occurrence. This process is
repeated until replicating the next pattern occurrence would surpass the
desired number of tasks.}

\subsection{\rev{WfSim}}
\label{sec:simulator}

An alternative to conducting scientific workflow research via real-world
experiments is to use simulation. Simulation is used in many computer
science domains and can address the limitations of real-world experiments.
In particular, real-world experiments are confined to those application
and platform configurations that are available to the researcher, and thus
typically can  only cover a small subset of the relevant scenarios that may be
encountered in practice. Furthermore, real-world experiments can be time-,
labor-, money-, and energy-intensive, as well as not perfectly reproducible.

\tool fosters the use of simulation for scientific workflow research,
e.g., the development of workflow scheduling and resource provisioning
algorithms, the development of workflow management systems, and the
evaluation of current an emerging computing platforms for workflow
executions. \rev{We do not
develop simulators as part of the \tool project. Instead, \tool' WfSim
component} catalogs open source WMS simulators (such as those
developed using the WRENCH framework~\cite{casanova-works-2018,
casanova2020fgcs}) that support the \rev{WfFormat workflow instance
format.} In other words, these simulators take as input workflow instances
\rev{in this format}  (either from actual workflow executions or synthetically
generated) and simulate their executions.  In the next section, we use two
of the simulators cataloged in \rev{WfSim} to quantify the extent to
which synthetic instances generated using \tool tools are
representative of real-world instances.

\subsection{\rev{WfCommons Python package}}
\label{sec:pyhton-package}

In order to allow users to analyze real workflow instances and to
generate synthetic workflow instances, the \tool framework provides a
collection of tools released as an open source Python
package~\cite{wfcommons-python, WfCommons-github}. Specifically, the
package leverages the SciPy ecosystem~\cite{virtanen2020scipy} for
performing probability distributions fitting to a series of data to
\rev{produce statistical summaries of workflow performance characteristics
(as described in Section~\ref{sec:wfchef})}.  In contrast to our previous
work~\cite{ferreiradasilva-escience-2014}, which used only two probability
distributions for generating workflow performance metrics, the
\tool Python package attempts to fit data with 23 probability
distributions provided as part of SciPy's statistics submodule. Workflow
recipes are instances of a class that defines methods for generating
synthetic workflow instances given a desired number of tasks.  The current
version of the \tool Python package\footnote{\rev{Version 0.6 was
released in May 2021.}} provides recipes for generating synthetic workflows
for all \rev{9} applications shown in Table~\ref{tab:instances}.  Detailed
documentation and examples can be found on the project's
website~\cite{wfcommons} and the online open access package
documentation~\cite{wfcommons-python}.

%% file: sec-experiments.tex
\section{Experimental Evaluation of Synthetic Generated Workflow Instances}
\label{sec:experiments}

\rev{In this section, we evaluate \tool and compare it to previously
proposed approaches. We first evaluate the realism of generated workflow
instances based on their structure (Section~\ref{sec:realism}), and then
based on their simulated execution using simulators of two state-of-the-art
WMSs (Section~\ref{sec:accuracy}).}

% subsection
\subsection{Experimental Scenarios}
\label{sec:scenarios}

We consider experimental scenarios defined by particular workflow instances
to be executed on particular platforms. To assess the \rev{realism and the
accuracy} of generated synthetic workflows, we have performed
real workflow executions with Pegasus \rev{and Makeflow}, and collected raw,
time-stamped event logs from these executions. These logs form
the ground truth to which we can compare simulated executions.

Actual workflow executions are conducted using the Chameleon Cloud
platform~\cite{keahey2020lessons}, an academic cloud testbed on which we
use homogeneous standard cloud units to run an HTCondor pool with shared
file system, a submit node (which runs Pegasus \rev{or Makeflow}), and a
data node placed in the WAN. \rev{We use 4 worker ``cloud units," where each
each cloud unit consists} of a 48-core 2.3GHz processor with 128 GiB of RAM.
The bandwidth between the submit node and worker nodes on these instances
is around 10Gbps.  Simulated workflow executions are obtained based on the
exact same hardware platform specification.

Whenever possible, for the experiments conducted in this section, we
contrast experimental results obtained with synthetic workflow
\rev{instances generated by \emph{\tool} to results obtained
using synthetic workflow instances generated by our previous work
(\emph{WorkflowHub})~\cite{da2020workflowhub}.
For the \emph{Epigenomics} and \emph{Montage} applications we also include
results obtained with an older, and very popular, framework called
\emph{WorkflowGenerator}~\cite{ferreiradasilva-escience-2014}.
We analyze a subset of 6 workflow applications, 4 from Pegasus and 2
from Makeflow (see Table~\ref{tab:applications}). We chose these applications
as they come from different science domains and have different
graph structures and/or computing requirements.}

\begin{table*}[!t]
  \centering
  \caption{\rev{Applications, number of tasks and systems used on our experiments.}}
  \label{tab:applications}
  \scriptsize
  \begin{tabular}{lll}
    \toprule
    Application & \multicolumn{1}{c}{Number of Tasks per Workflow Instance} & \multicolumn{1}{c}{WMS} \\
    \midrule
    Blast       & [45, 105, 305] & Makeflow \\
    BWA         & [106, 1006] & Makeflow \\
    Cycles      & [69, 135, 136, 203, 221, 268, 333, 401, 439, 440, 659, 663, 664, 876,  995, 1093, 1313, 1324, 1985, 2183, 2184, 3275, 4364, 6545] & Pegasus \\
    Epigenomics & [43, 75, 121, 127, 225, 235, 243, 265, 349, 407, 423, 447, 509, 517, 561, 579, 673, 715, 795, 819, 865, 985, 1097, 1123, 1399, 1697] & Pegasus \\
    1000Genome  & [54, 84, 106, 158, 166, 210, 248, 262, 314, 330, 366, 412, 418, 470, 494, 522, 574, 576, 658, 740, 822, 904] & Pegasus \\
    Montage     & [180, 312, 474, 621, 621, 750, 1068, 1314, 1740, 2124, 4848, 6450, 7119, 9807] & Pegasus \\
    \bottomrule
  \end{tabular}
\end{table*}

% subsection
\subsection{\rev{Evaluating the Realism of Synthetic Workflow Instances}}
\label{sec:realism}

\rev{To evaluate the structure of the generated synthetic workflow
instances, we developed a metric called \textbf{Type Hash Frequency (THF)}.
The THF metric is the Root Mean Square Error (RMSE) between the frequency
of task type hashes for a synthetic workflow instance and that for the
real workflow instance with the same number of tasks. Recall from
Section~\ref{sec:wfchef} that a task's type hash encodes information about
the task's type (kind of computation), but also about the task's ancestors
and descendants.  Therefore, the lower the THF of a synthetic workflow instance,
the more similar it is to the real workflow instances.
}

\begin{figure*}[!t]
  \centering
  \begin{subfigure}[t]{0.45\linewidth}
    \centering
    \includegraphics[width=\linewidth]{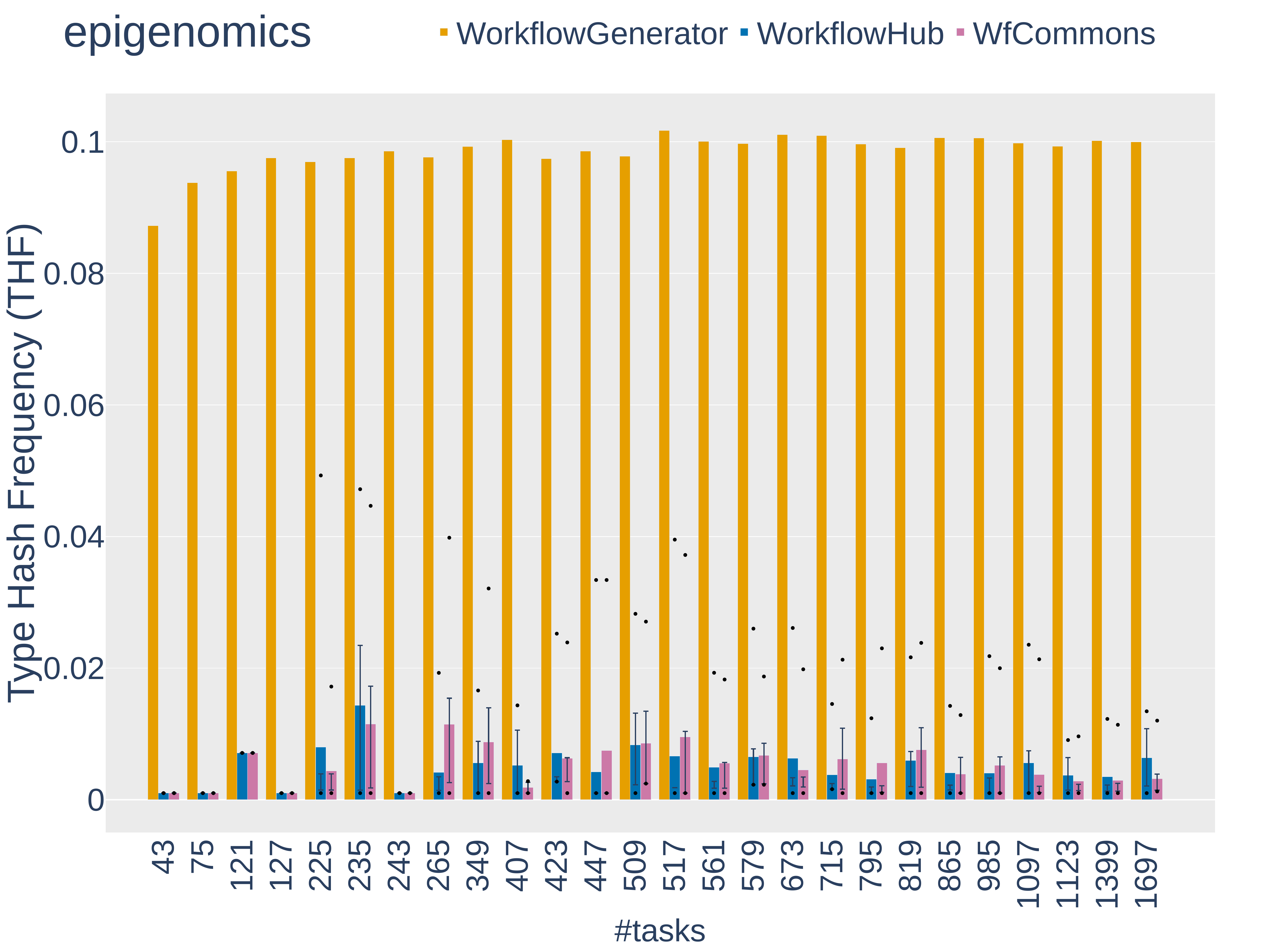}
    \vspace{-15pt}
    \caption{Epigenomics}
    \label{fig:epigenomicsTHF}
  \end{subfigure}
  \quad
  \begin{subfigure}[t]{0.45\linewidth}
    \centering
    \includegraphics[width=\linewidth]{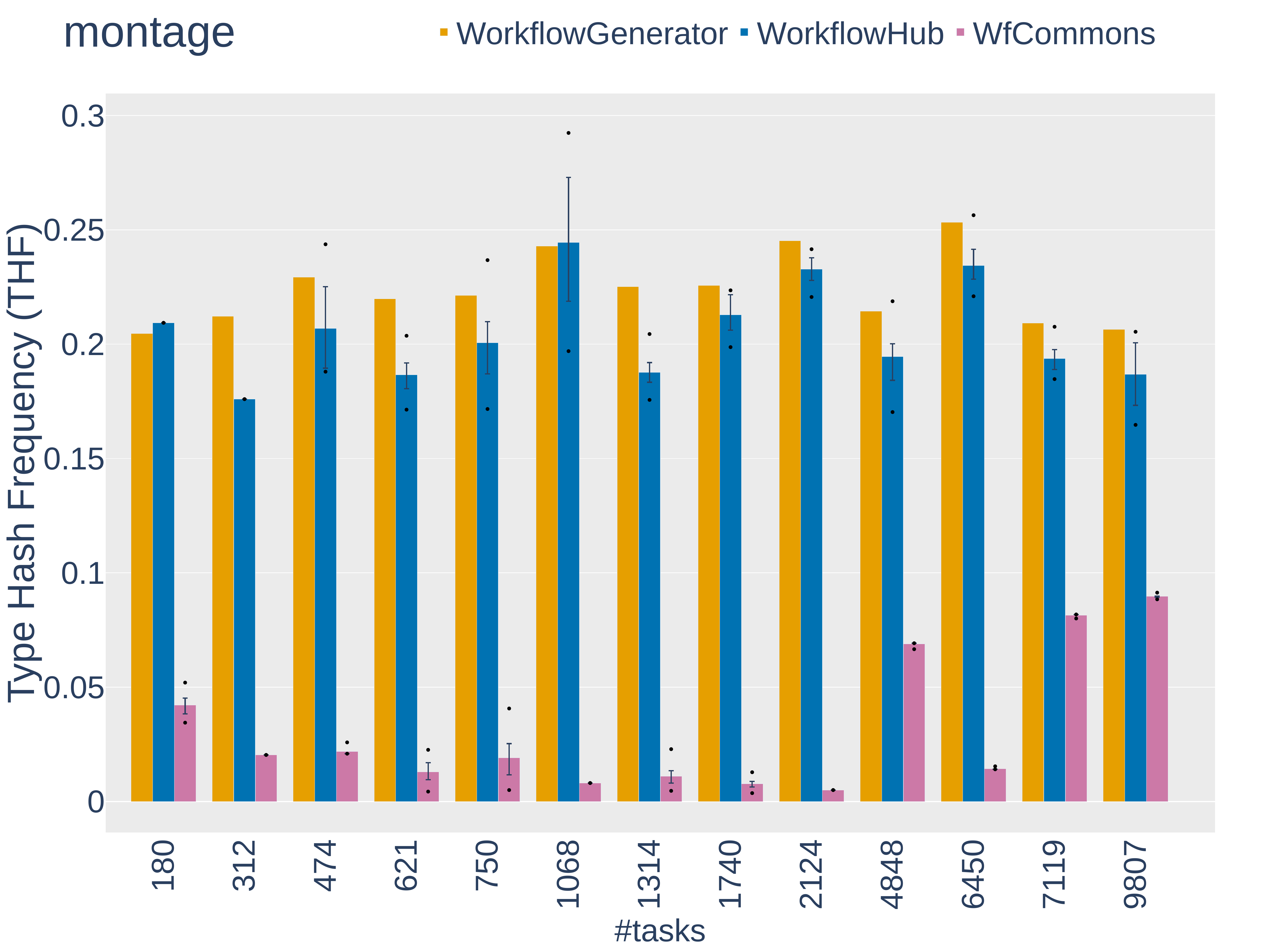}
    \vspace{-15pt}
    \caption{Montage}
    \label{fig:montageTHF}
  \end{subfigure}
  \quad
  \begin{subfigure}[t]{0.45\linewidth}
    \centering
    \includegraphics[width=\linewidth]{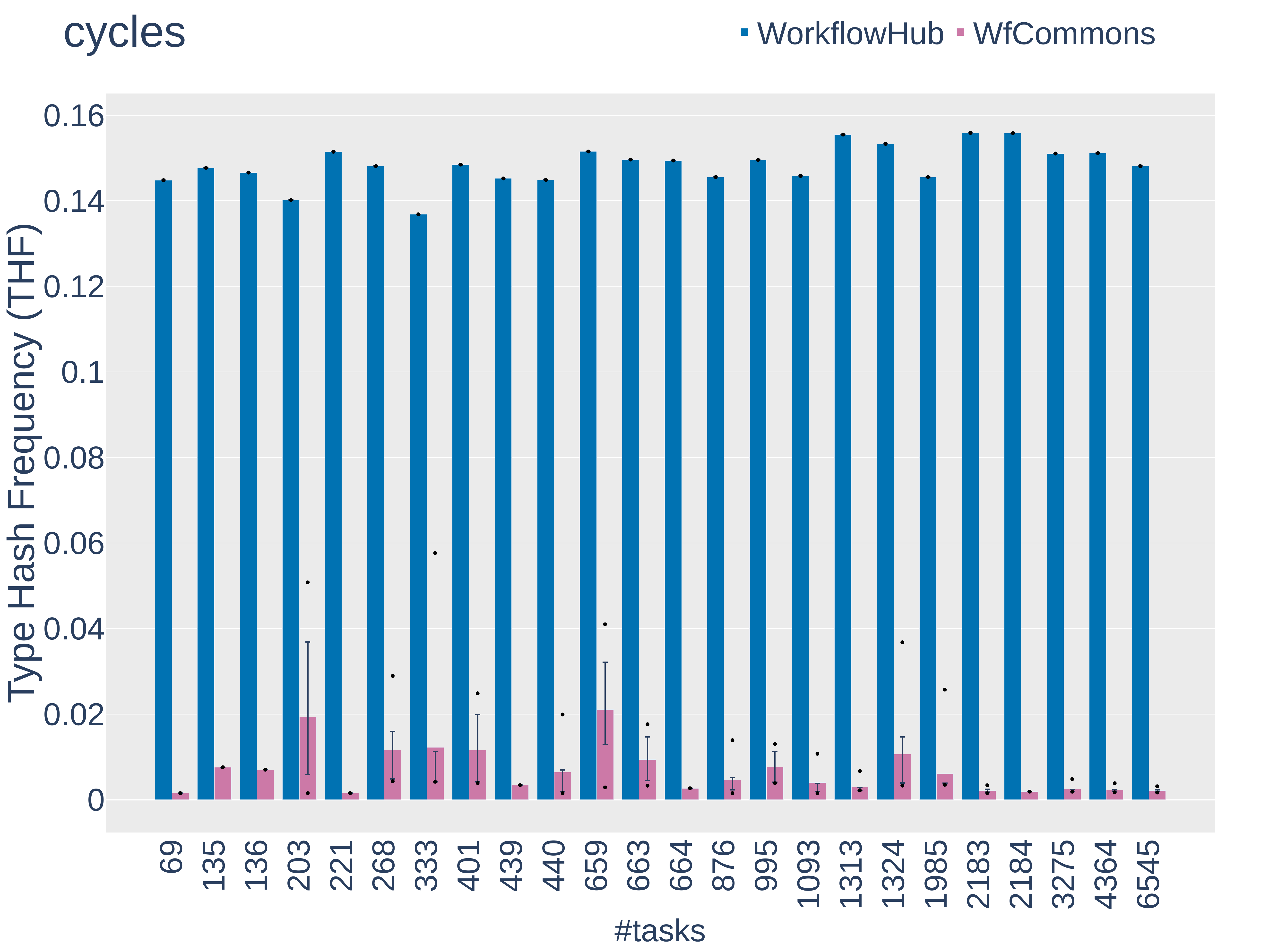}
    \vspace{-15pt}
    \caption{Cycles}
    \label{fig:cyclesTHF}
  \end{subfigure}
  \quad
  \begin{subfigure}[t]{0.45\linewidth}
    \centering
    \includegraphics[width=\linewidth]{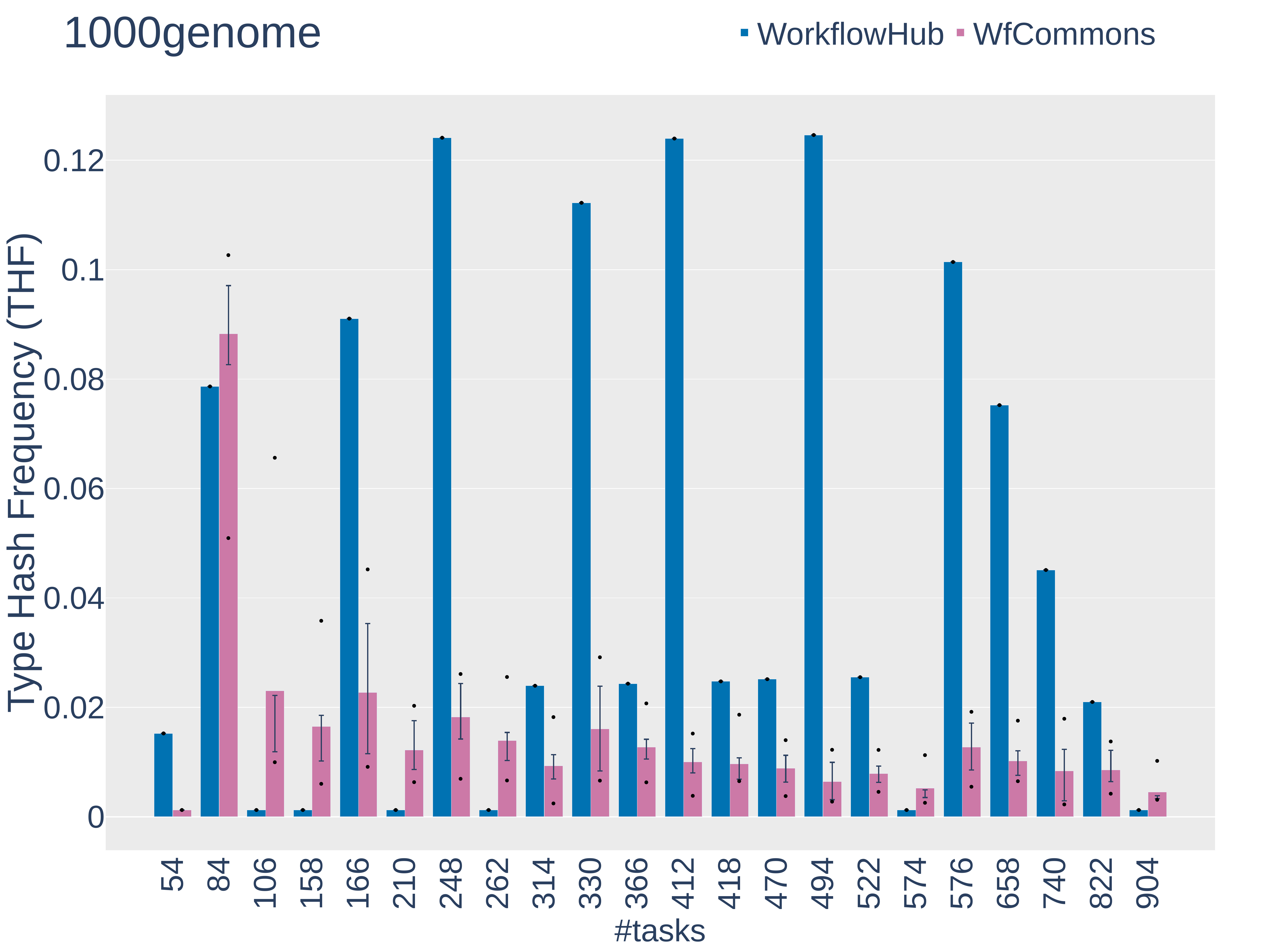}
    \vspace{-15pt}
    \caption{1000Genome}
    \label{fig:genomeTHF}
  \end{subfigure}
  \quad
  \begin{subfigure}[t]{0.45\linewidth}
    \centering
    \includegraphics[width=\linewidth]{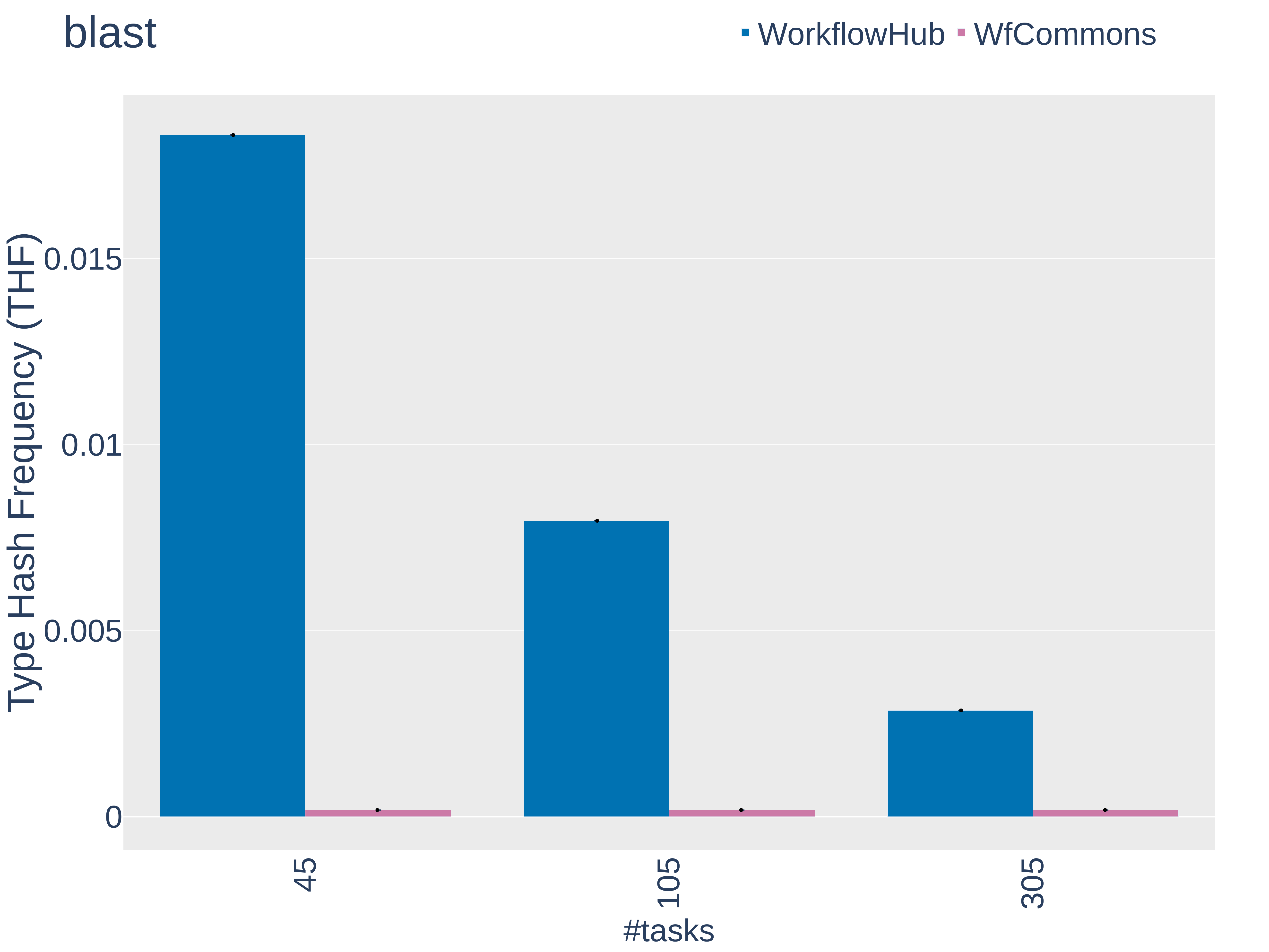}
    \vspace{-15pt}
    \caption{Blast}
    \label{fig:blastTHF}
  \end{subfigure}
  \quad
  \begin{subfigure}[t]{0.45\linewidth}
    \centering
    \includegraphics[width=\linewidth]{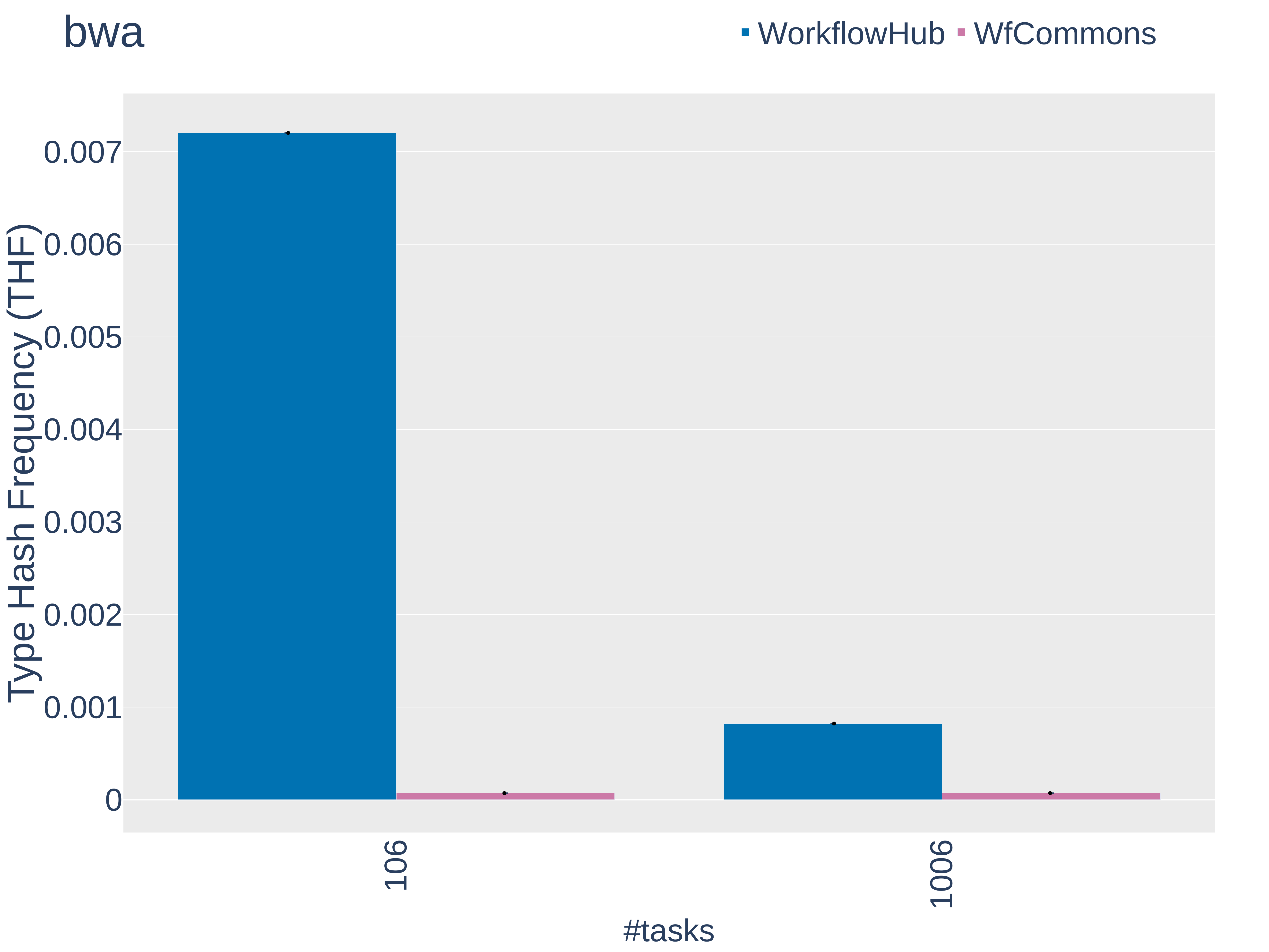}
    \vspace{-15pt}
    \caption{BWA}
    \label{fig:bwaTHF}
  \end{subfigure}

  \caption{\rev{THF of synthetic workflow instances. Bar heights are average values.
  Error bars show the range between the third quartile (Q3) and the first
  quartile (Q1), and minimum and maximum values as black dots.}}
  \label{fig:thf}
\end{figure*}

For each real workflow instance of each selected application, as archived
on \tool, we use \tool' Python package for generating a synthetic
workflow instance with the same number of tasks.  For comparison purposes,
we also generate synthetic instances using the generators from our previous
works.

\rev{Figure~\ref{fig:thf} shows THF results for all 6 applications, for
\tool, WorkflowHub, and WorkflowGenerator. WorkflowGenerator only supports
2 of these application, which is why it is
not included in all 6 plots.  \tool and WorkflowHub use randomization
in their heuristics, thus for each number of tasks we generate 10 sample
synthetic workflows with each tool. The heights of the error bars in
Figure~\ref{fig:thf} show average THF values, and error bars show the range
between the third quartile (Q3) and the first quartile (Q1), in which 50\%
of the results lie.  Error bars also show minimum and maximum values. Note
that error bars are of zero length for the Blast
(Figure~\ref{fig:blastTHF}) and BWA (Figure~\ref{fig:bwaTHF}) applications.
This is because workflows for these applications comprise a simple graph
structure: there is only one task that can be replicated to produce
synthetic workflow instances of different sizes. As a result, both WorkflowHub
and \tool each produce ten identical synthetic workflow instances.}

\rev{Overall, \tool yields the lowest THF values in most cases, often
achieving low values in the absolute sense, meaning that the synthetic
workflow instances it generates are representative of real workflow instances.
Synthetic workflow instances produced by WorkflowGenerator have fixed graph
structures, thus scaling up (resp. down) the number of tasks is simply done
by replicating (resp. pruning) predefined subgraphs of the workflow. As a
result, the generated workflow instances do not capture distinct patterns
of the workflow graph produced by different sets of input data/parameters.
For instance, for the Epigenomics workflow
(Figure~\ref{fig:epigenomicsTHF}) smaller instances are composed of a
single or few chains of tasks, while larger instances are composed of
several chains but also multiple branches (that can be composed of
different numbers of chains). WorkflowGenerator is unable to capture this
pattern.  Synthetic workflow instances generated by the manually-crafted
WorkflowHub generators are on average about 52\% more realistic when
compared to WorkflowGenerator. However, WorkflowHub still does not entirely
capture all workflow patterns. For the Montage workflow
(Figure~\ref{fig:montageTHF}), real-world instances are obtained using two
different image datasets (2MASS and DSS)~\cite{deelman-fgcs-2015}.
Although these workflow instances are composed of the same set of
executables, their graph structures differ significantly. WorkflowHub
attempts to find a single structure to capture both cases, while
\tool can precisely identify both distinct patterns.  Similar results are
observed for Cycles and 1000Genome (Figures~\ref{fig:cyclesTHF}
and~\ref{fig:genomeTHF}).  For Blast and BWA
(Figures~\ref{fig:blastTHF} and~\ref{fig:bwaTHF}), THF values are very low,
(but still with \tool leading to the best results) due to the simple
structure of these workflows.}

% subsection
\subsection{Evaluating the Accuracy of Synthetic Workflow Instances}
\label{sec:accuracy}

We use simulators of two state-of-the-art WMSs,
Pegasus~\cite{deelman-fgcs-2015} \rev{and
Makeflow~\cite{albrecht2012makeflow}}, as a case study for evaluation and
validation purposes. These simulators are described
in~\cite{casanova2020fgcs} \rev{(note that the Makeflow simulator is really a
simulator of WorkQueue, an execution engine used by Makeflow)}.  \rev{Both
Pegasus and Makeflow are being used in production to execute workflows for
dozens of high-profile applications in a wide range of scientific domains
and on a wide range of platforms.  We used both systems} to execute
workflows on a cloud environment for the purpose of collecting execution
logs for building real workflow instances, as described in
Section~\ref{sec:instances}.  The simulators are built using
WRENCH~\cite{casanova-works-2018, casanova2020fgcs}, a framework for
implementing simulators of WMSs that are accurate and can run scalably on a
single computer, while requiring minimal software development effort.  The
work in~\cite{casanova2020fgcs} demonstrates that WRENCH achieves these
objectives, and provides high simulation accuracy for workflow executions
using both \rev{Pegagus and Makeflow}.

In~\cite{casanova2020fgcs, da2020workflowhub}, we have already demonstrated
that the simulation framework used in our previous set of
tools~\cite{ferreiradasilva-escience-2014}
suffers from significant discrepancies from actual executions.  These
discrepancies mostly stem from the use of a simplistic network simulation
model, and from the simulator not capturing relevant details of the system,
and thus of the workflow execution.  Therefore, to reach fair conclusions
regarding the validity of synthetic workflow instances, in this paper we
only use the more accurate WRENCH-based simulators for all experiments.
Using these simulators, we quantify the extent to which the simulated
execution of generated synthetic workflow instances (generated using our
previous work and using \tool) is similar to that of real workflow
instances.  The simulator implementations, details on the calibration
procedure, and experimental scenarios used in the rest of this section are
all publicly available online~\cite{wrench-pegasus, wrench-workqueue}.

\rev{We perform experiments using simulators for the same subset of
workflow applications}. For each application, we run the simulator for a
reference real workflow instance and for synthetic instances. \rev{The goal
is to quantify the discrepancies between the simulated execution of a
synthetic workflow instance and that of a real workflow instance with the
same number of tasks, using the absolute relative difference between the
simulated makespans (i.e., overall execution times in seconds). This metric
is commonly used in the literature to quantify simulation error.}

\begin{figure*}[!t]
  \centering
  \begin{subfigure}[t]{0.45\linewidth}
    \centering
    \includegraphics[width=\linewidth]{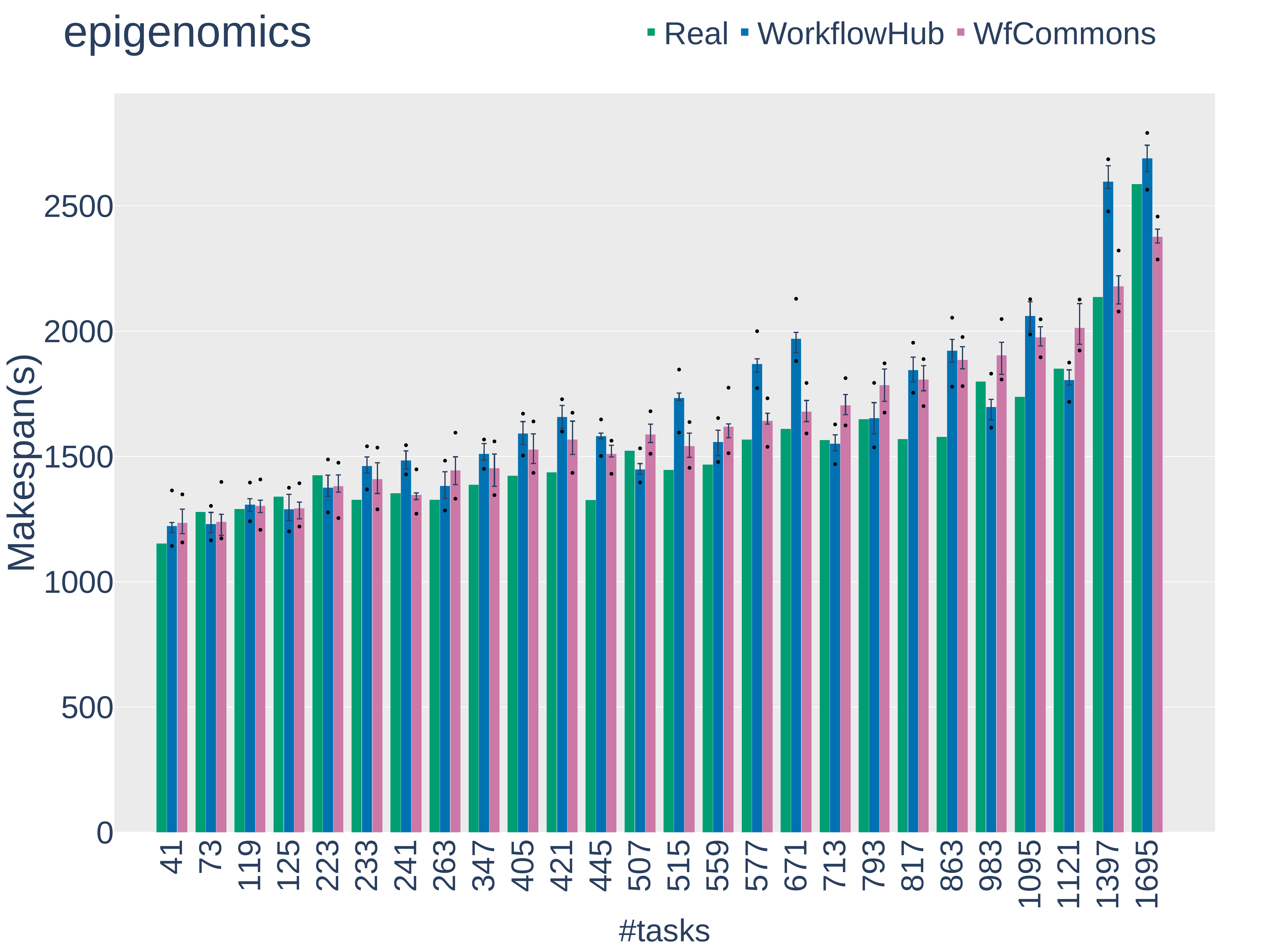}
    \caption{Egipenomics}
    \label{fig:epigenomics_makespan}
  \end{subfigure}
  \quad
  \begin{subfigure}[t]{0.45\linewidth}
    \centering
    \includegraphics[width=\linewidth]{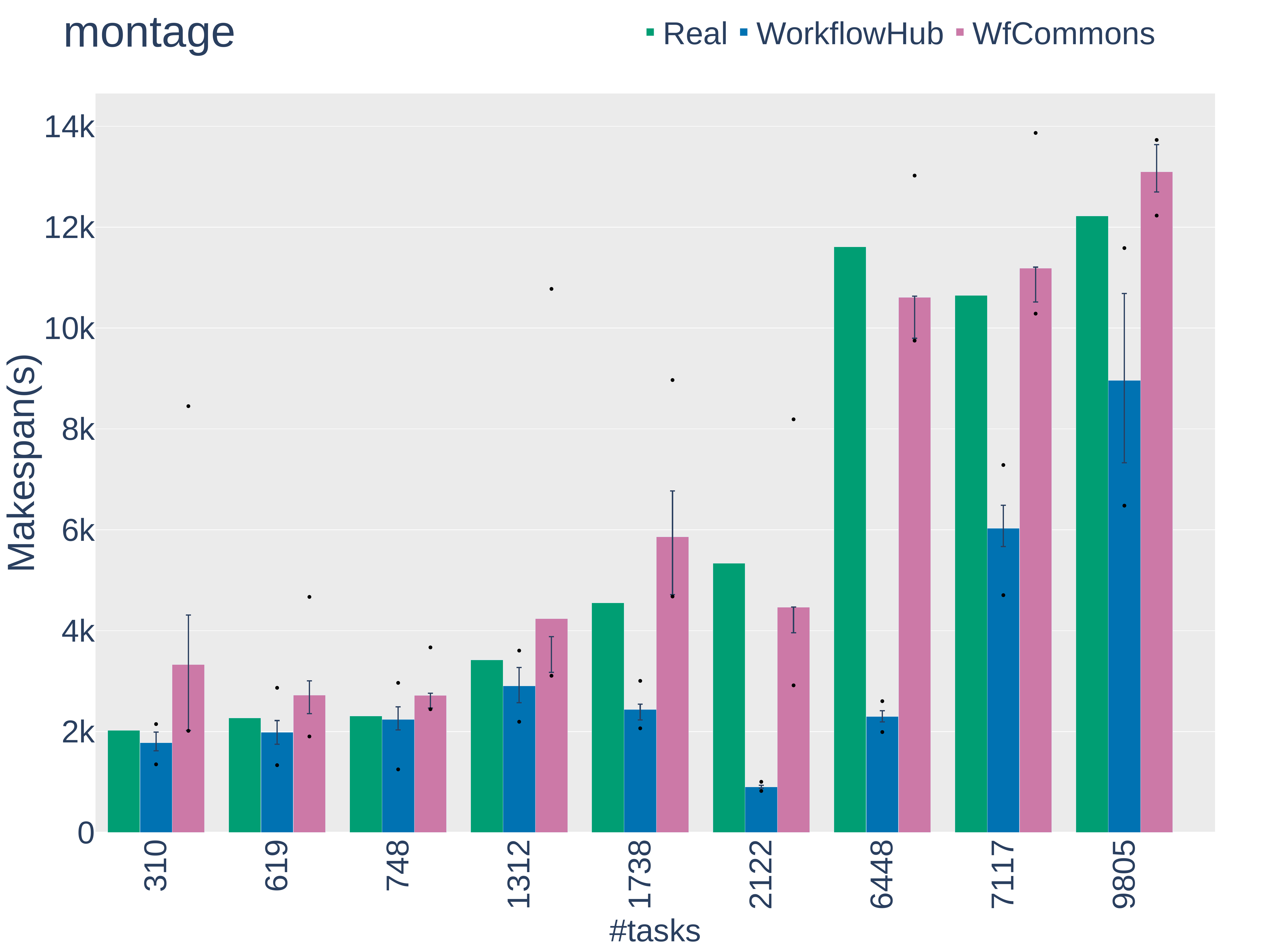}
    \caption{Montage}
    \label{fig:montage_makespan}
  \end{subfigure}
  \quad
  \begin{subfigure}[t]{0.45\linewidth}
    \centering
    \includegraphics[width=\linewidth]{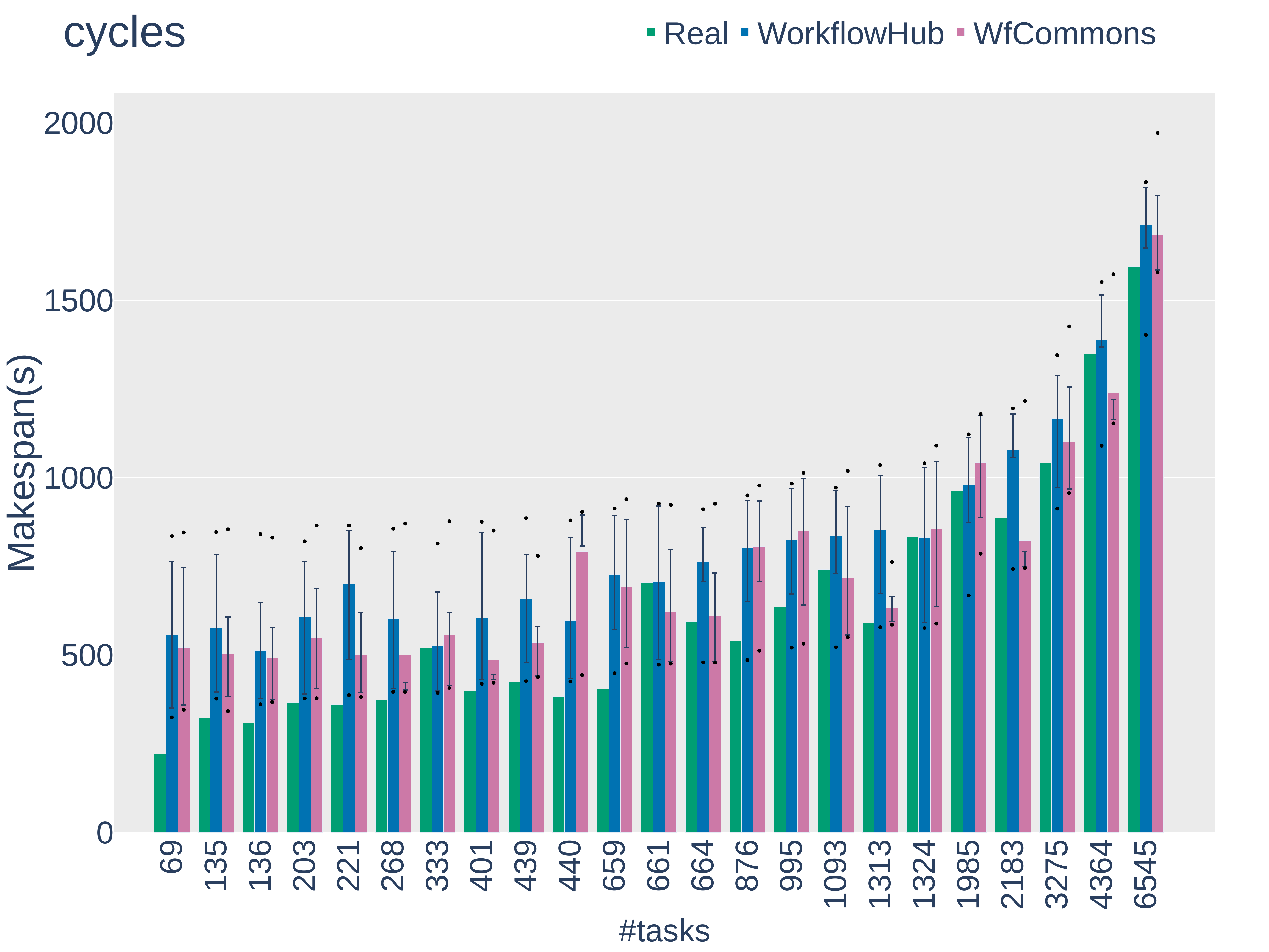}
    \caption{Cycles}
    \label{fig:cycles_makespan}
  \end{subfigure}
  \quad
  \begin{subfigure}[t]{0.45\linewidth}
    \centering
    \includegraphics[width=\linewidth]{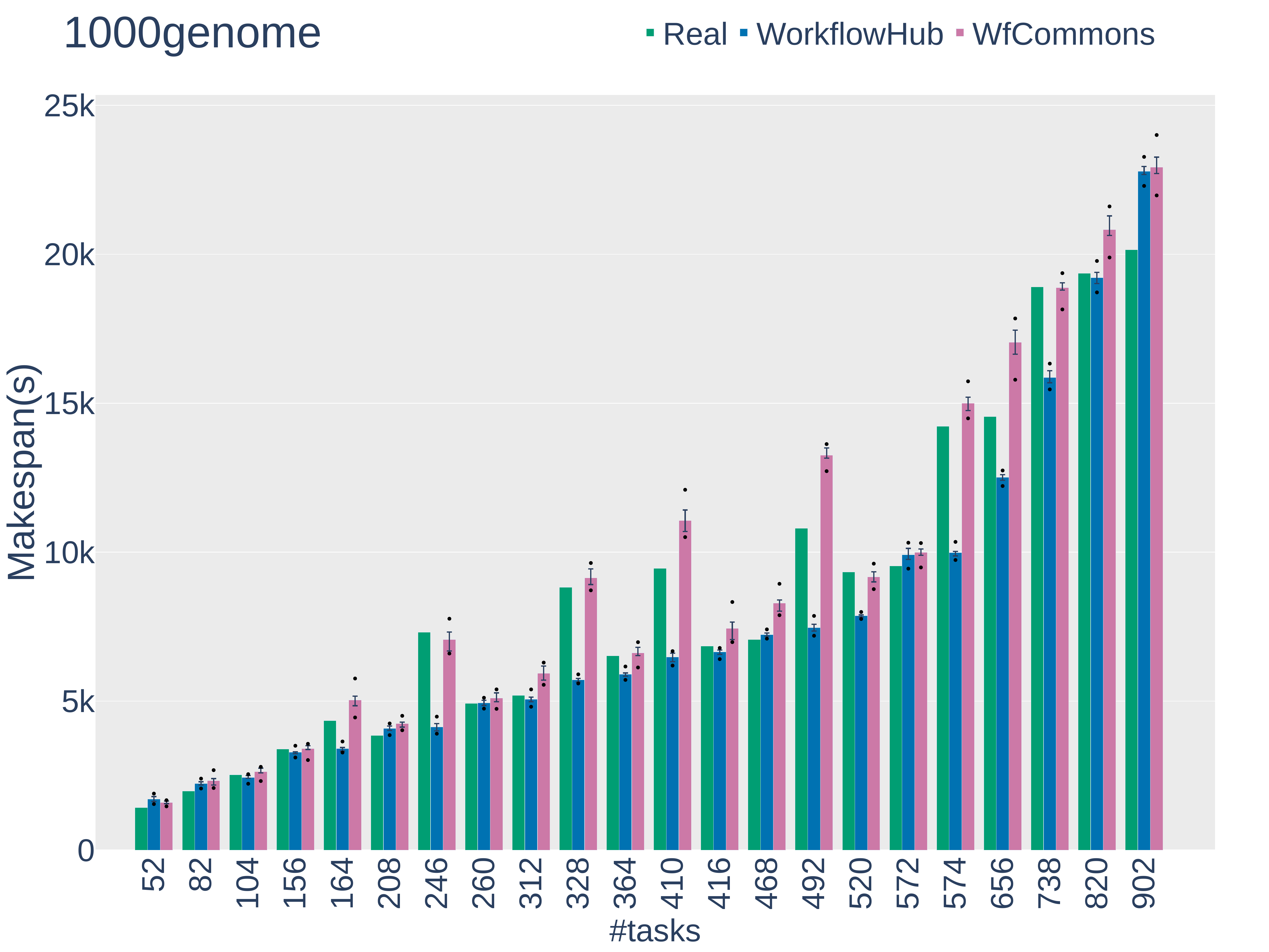}
    \caption{1000Genome}
    \label{fig:genome_makespan}
  \end{subfigure}
  \quad
  \begin{subfigure}[t]{0.45\linewidth}
    \centering
    \includegraphics[width=\linewidth]{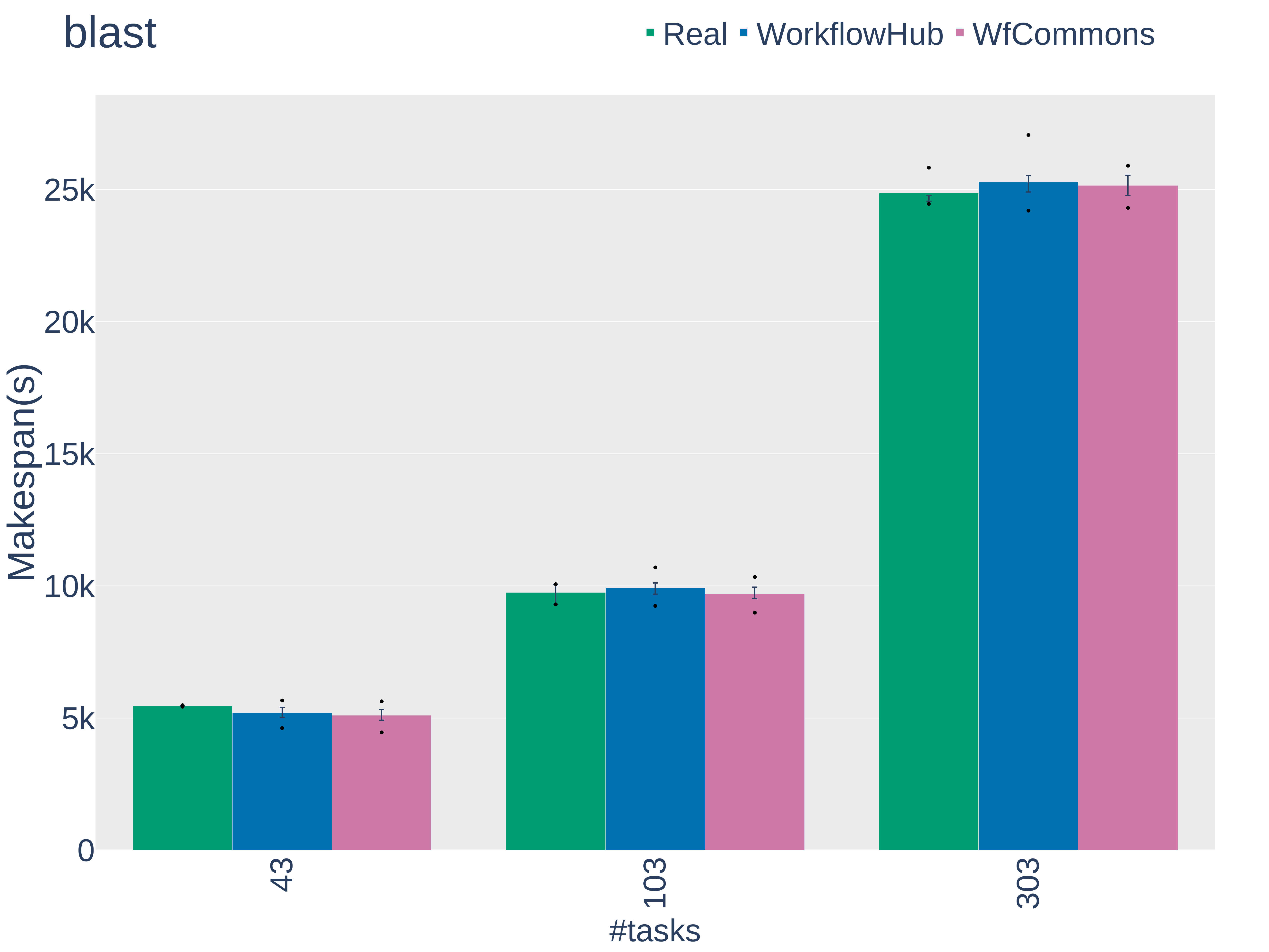}
    \caption{Blast}
    \label{fig:blast_makespan}
  \end{subfigure}
  \quad
  \begin{subfigure}[t]{0.45\linewidth}
    \centering
    \includegraphics[width=\linewidth]{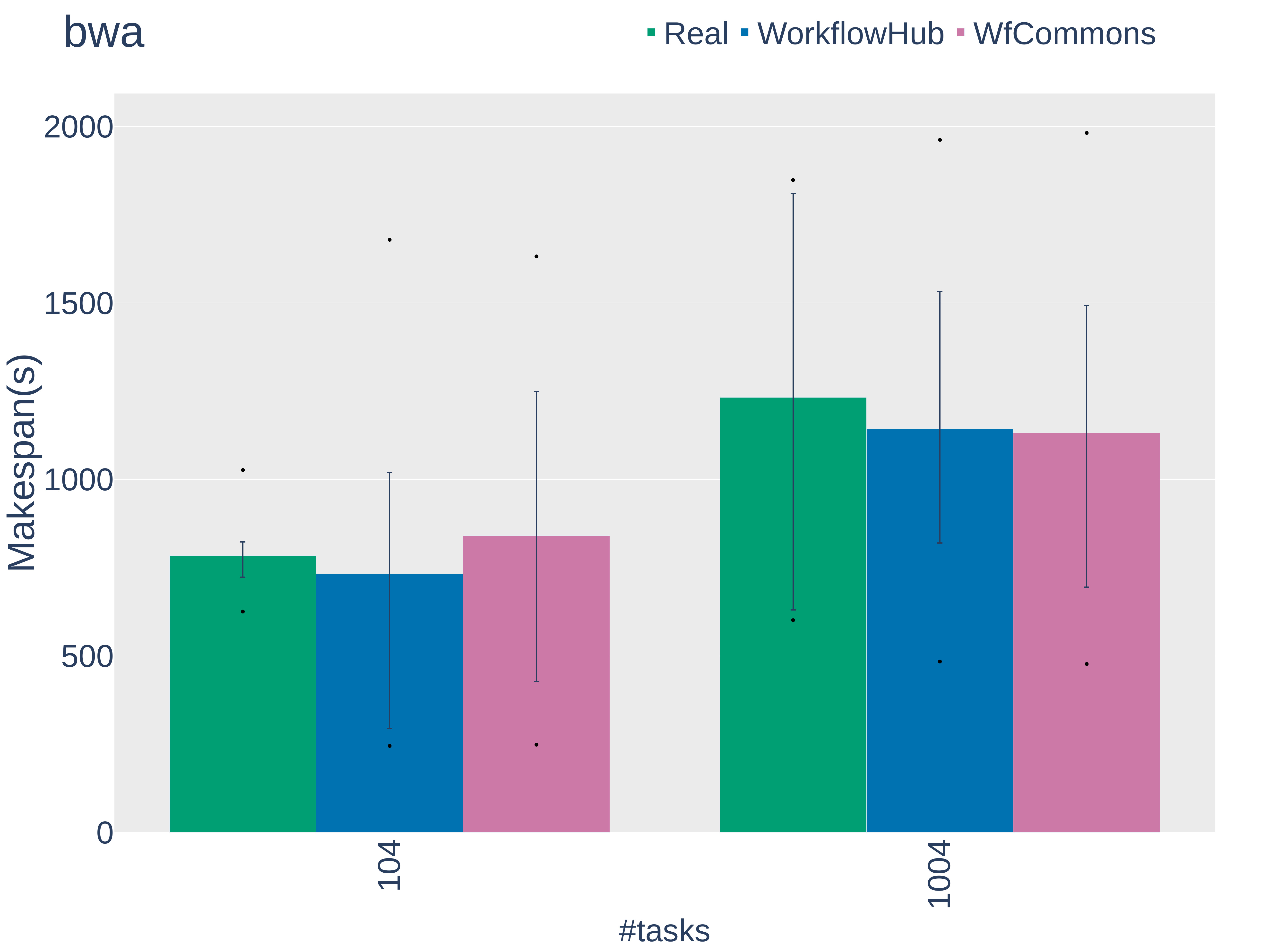}
    \caption{BWA}
    \label{fig:bwa_makespan}
  \end{subfigure}
  \caption{\rev{Simulated makespan for workflow instances. Bar heights
  represent average values. Error bars show the range between the third
  quartile (Q3) and the first quartile (Q1), and minimum and maximum
  values as black dots.}}
  \label{fig:makespan}
\end{figure*}

\rev{Figure~\ref{fig:makespan} shows simulated makespans
for all 6 applications for \tool, WorkflowHub, and the real (Real)
instances. In this section, we omit all results for WorkflowGenerator
as it performs very poorly, as is expected given the results in
the previous section.
Recall that \tool and
WorkflowHub have randomization in their heuristics, therefore the values
shown in Figure~\ref{fig:makespan} are averages computed over 10 sample
generated synthetic workflows for each tool and application. Note that error
bars are also shown for ``Real" executions of Blast
(Figure~\ref{fig:blast_makespan}) and BWA (Figure~\ref{fig:bwa_makespan})
as \tool archives more than one workflow instance for a given number of tasks
for these two applications.}

Overall, we observe larger discrepancy between simulated executions of
synthetic workflow instances and that of real workflow instances when the
synthetic instances are generated by WorkflowHub.  By contrast, \tool'
generation method produce workflow instances whose executions are more
closely matched to that of  real workflow instances (even though some
discrepancies necessarily remain due to random sampling effects).  \rev{For
Epigenomics (Figure~\ref{fig:epigenomics_makespan}), both frameworks lead
to many similar average makespan values, but WorkflowHub leads to
significantly higher (and less realistic) makespans for instances comprised
of 515, 577, 671, and 1397 tasks. By manually inspecting these generated synthetic
instances, we observe a reduction in the number of branches in the workflow
and an increase in the number of tasks in chains of tasks. As a result, the
parallelism of the workflow is decreased, which explains the longer makespans.  In
the real instances, this parallelism reduction does not occur.
\tool-generated instances are more representative of real instances but
some discrepancies still occur for instances comprised of 713, 793, and
1695 tasks, albeit with smaller magnitude.
Results for Montage (Figure~\ref{fig:montage_makespan}) corroborate
the findings presented in Section~\ref{sec:realism}, with a large
advantage for \tool over WorkflowHub.  Results for Cycles
(Figure~\ref{fig:cycles_makespan}) are similar, with \tool instances
leading to makespans more in line with that of real instances for most cases.
Simulated results for 1000Genome instances
(Figure~\ref{fig:genome_makespan}) are similar to Epigenomics results, in that
WorkflowHub leads to very inaccurate makespans for some instances. These
large discrepancies occur for particular numbers of tasks (246,
328, 410, 492, 574, 656, and 738), which correspond to cases in which
a new chromosome is added.  WorkflowHub is not able to capture these
changes in the workflow structures. By contrast, \tool leads to more
accurate results, albeit with some remaining discrepancies.
Finally, results for Blast
(Figure~\ref{fig:blast_makespan}) and BWA (Figure~\ref{fig:bwa_makespan}) show
that both WorkflowHub and \tool lead to accurate makespans, which is
expected given the simple structure of these workflows. }

%% file: sec-energy.tex
\section{\rev{Case Study: Estimating Energy Consumption of Large-Scale Workflows}}
\label{sec:energy}

\rev{Energy-efficient computing has received much attention in the past few
years.  With the advance of computing capabilities, applications become
more complex and consume more resources, thus leading to increased energy
usage~\cite{zakarya2018energy}.  While the development of fully renewable
computing facilities is on the rise, there is still a pressing need to
reduce the power consumption of computation, which in turn reduces its
carbon footprint.  In the context of scientific workflows, several works
have proposed solutions to optimize workflow executions while respecting
energy consumption constraints~\cite{orgerie2014survey}.}

\rev{In~\cite{ferreiradasilva-iccs-2019, ferreiradasilva2020jocs}, we have
proposed and validated a power consumption model that accounts for CPU
utilization, computations that execute on multi-core compute nodes, and I/O
operations (including the idle power consumption caused by waiting for
these operations to complete). In this work, we leverage this model, to
estimate the energy consumption of the execution of large-scale workflows.
We use Montage workflows as a case-study and compute the estimated energy
consumption of the execution of real workflow instances using our model. We
then compute the estimated energy consumption of the execution of synthetic
Montage instances generated by \tool. We generate synthetic instances with
approximately the same number of tasks as their real counterparts, so as to
validate the accuracy of the generated instances. Additionally, we generate
larger instances of that available real Montage workflow instances, that
is, with 10K, 25K, 50K, 75K, 100K, 150K, 200K, and 250K tasks. The goal is
to demonstrate the usefulness of \tool in assessing the energy consumption
of Montage applications at scales for which data from actual executions is
not available.}

\begin{figure}[!t]
  \centering
  \includegraphics[width=\linewidth]{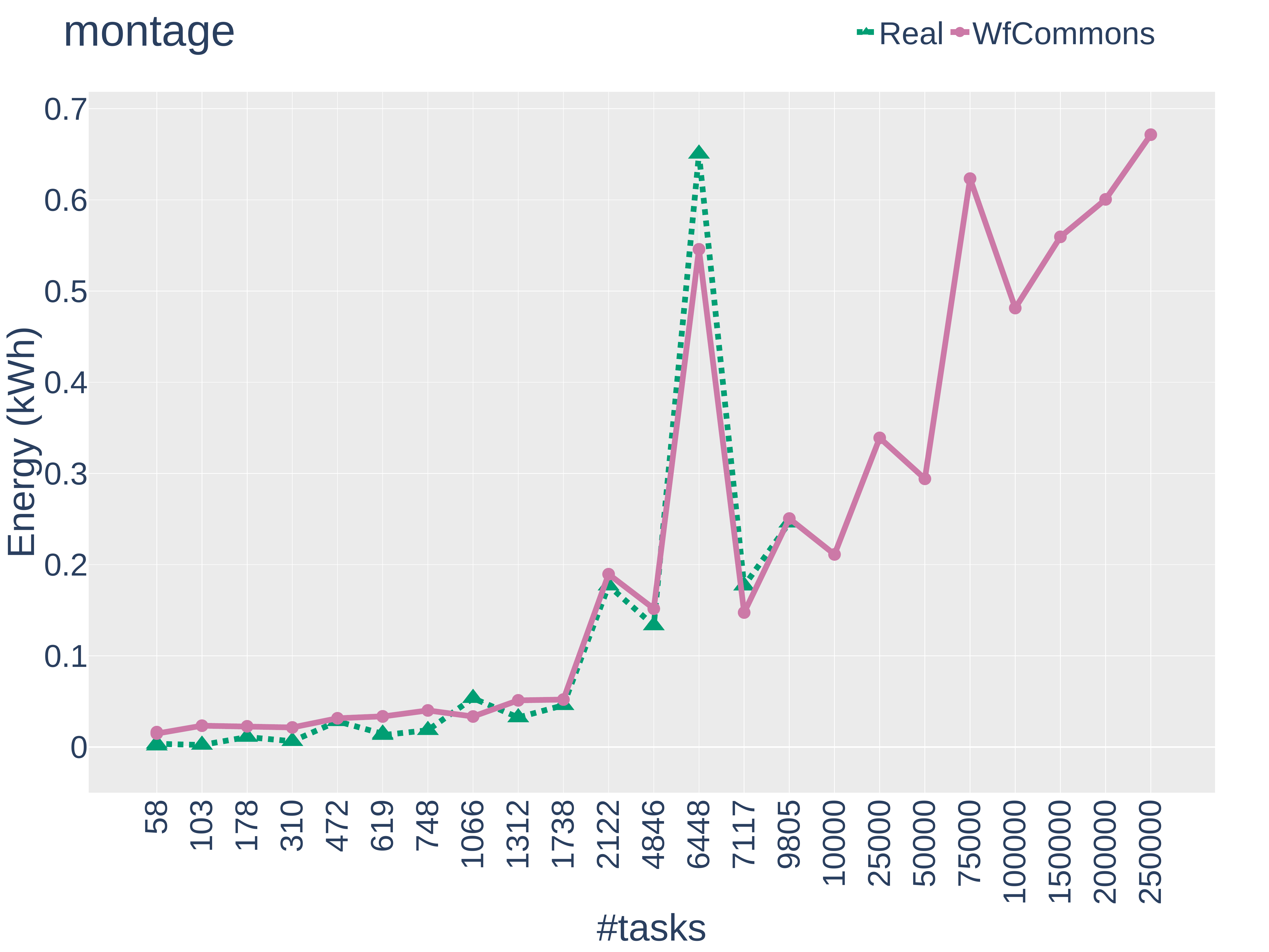}
    \caption{\rev{Estimated energy consumption (in kWh) of the (simulated) executions of
           synthetic and real Montage workflow instances. Synthetics instances are
           generated at scales beyond that of the available real instances.}}
  \label{fig:energy}
\end{figure}

\rev{Figure~\ref{fig:energy} shows simulated energy consumption vs.  number
of workflow tasks, for both real and synthetic Montage workflow instances.
Simulated executions are for the same hardware platform specification as that
described in Section~\ref{sec:scenarios}.
Intriguingly, energy consumption for real instances does not monotonically
increase with the number of workflow tasks. Similar non-monotonic behavior is
observed for all other 5 applications considered in this work (results not
shown).  For Montage instances that comprise 2,212 and 6,448 tasks we note
large energy consumption spikes.  Manual inspection of these instances
reveals that the number of tasks in the fan-out portions of the workflow
graph significantly diminishes when compared to the adjacent instances
(which have fewer branches and stretched fan-out patterns). This
application-specific feature leads to a reduction of the number of tasks
that are ready for execution at particular points in time, thus causes the WMS
(or rather the scheduling algorithm it employs) to underutilize the
resources. This lowered resource utilization results in higher makespans
(as seen in Figure~\ref{fig:montage_makespan}). It also results in a large
increase in energy consumption due to the inherent energy consumption of
having machines in operation, even if idle (i.e., due to \emph{static
power consumption}).}

\rev{The key results here is that synthetic workflow instances generated by
\tool produce very similar energy consumption profiles to that of the
result instances.  More importantly, the spikes observed for the real
instances are also observed for synthetic instances. This shows that \tool
is able to account for application-specific features and patterns
accurately when producing workflow recipes.  Figure~\ref{fig:energy} shows
8 data points for synthetic instances that go beyond the scale of available
real instances. These data points show a large energy-consumption spike for
the instance with 75K tasks. Given the accuracy of the energy consumption
estimate with synthetic instances for scales up to 9,805 tasks, there is
good confidence that a real Montage execution with 75K tasks would also
experience a large energy consumption spike.}

\rev{
The overall conclusion from the above results, which corroborates results in
previous sections, is that \tool generates
accurate (both at structural and performance metrics level) synthetic
workflow instances. These instances can be used to study workflow execution
behavior, as demonstrated for energy consumption behavior in this case
study,  at scales beyond that for which real execution data is available.
}

%% file: sec-conclusion.tex
\section{Conclusion}
\label{sec:conclusion}

In this paper, we have presented the \tool project, a community
framework for constructing and archiving workflow instances,
analyzing these instances, producing realistic synthetic workflow
instances, and simulating workflow executions using these
instances. \tool provides a collection of tools for
constructing ``workflow recipes" based on instances collected from
the real-world execution of workflow applications.  These workflow recipes
can then be used to produce synthetic workflow instances.  These
synthetic instances that can enable a variety of novel workflow systems
research and development activities.   We have demonstrated experimentally
that the synthetic instances generated by \tool are realistic,
much more so than those produced by previously available generators. More
specifically, \tool is able to generate synthetic workflow
instances at various scales while preserving key application-specific
\rev{structural patterns and performance characteristics. We have showcased
the usefulness of \tool via a case study focused on the energy consumption
of workflow executions. Specifically, we have shown that using the
generated synthetic workflow instances lead to experimental results that
are in line with that obtained with real workflow instances, while making
it possible to explore scenarios for workflow scales beyond that of
available real workflow instances.}

\tool is open-source and welcomes contributors. It currently provides
a collection of \rev{165 workflow instances} derived from actual executions, and can
generate synthetic workflows from \rev{9} applications from 4 science domains.
\rev{Version 0.6 was released in May 2021}. We refer the reader to
\url{https://wfcommons.org} for software, documentation, and links to
collections of instances and simulators.

\rev{A short-term future work direction is the development of additional
execution logs parsers for state-of-the-art workflow
systems~\cite{workflow-systems}.  These parsers will enable \tool to}
provide continuous automated development of novel workflow recipes to
broaden the number of science domains in which \tool can potentially impact
research and development efforts.  Another future work direction is the use
of synthetic workflow instances to support workflow-focused education and
training,  e.g., for designing simulation-driven activities in which
students acquire knowledge by experimenting with various workflow
scenarios~\cite{tanaka2019eduhpc}.